\documentclass{aa}  
\usepackage{psfig}
\usepackage{graphics}
\usepackage{epsfig}
\usepackage{amssymb}

\begin{document}
\title{A new sample of giant radio galaxies from the WENSS survey}
\subtitle{I - Sample definition, selection effects and first results}
\author{A.P. Schoenmakers\inst{1,2,3}\thanks{{\it Present address: }ASTRON, P.O. Box 2, 7990~AA Dwingeloo, The Netherlands}
\and A.G. de Bruyn\inst{3,4}
\and H.J.A. R\"{o}ttgering\inst{2}
\and H. van der Laan\inst{1}} 
\institute{Astronomical Institute, Utrecht University, P.O. Box 80\,000,
3508~TA Utrecht, The Netherlands 
\and {Sterrewacht Leiden, Leiden University, P.O. Box 9513, 2300~RA Leiden, The Netherlands}
\and {ASTRON, P.O. Box 2, 7990~AA Dwingeloo, The Netherlands}
\and {Kapteyn Astronomical Institute, University of Groningen, P.O. Box 800, 9700~AV Groningen, The Netherlands}}
\offprints{A.P. Schoenmakers (Schoenmakers@astron.nl)}
\date{Received ; accepted}
\titlerunning{A new sample of GRGs}
\authorrunning{A.P. Schoenmakers et al.}
\abstract{
We have used the Westerbork Northern Sky Survey (WENSS) to define a
sample of 47 low redshift ($z \la 0.4$) giant radio galaxies.
This is the largest sample yet of such radio sources originating from
a single survey.  We present radio maps of the newly discovered giants
and optical images and spectra of their host galaxies.  We present
some of their properties and discuss the selection effects.  We
investigate the distribution of the sources in the radio power --
linear size ($P-D$) diagram, and how these parameters relate to the
redshifts of the sources in our sample.  We find 
a strong drop in the number of sources with a linear size above 2
Mpc. We argue that this is not a result of selection effects, but that
it indicates either a strong luminosity evolution of radio sources of
such a size, or that a large fraction
of radio sources `switch off' before they are able to grow to a size of
2 Mpc. 
\keywords{
Galaxies: active -- Intergalactic medium -- Galaxies: jets -- Radio continuum: galaxies
}}

\maketitle

\section{Introduction}

The central activity in a fraction of Active
Galactic Nuclei (AGN) is capable of producing relativistic outflows of
matter, the so-called `jets', for a prolonged period of time, possibly
up to a few $10^8$ yr.  These jets, when powerful enough, inflate a
cocoon (e.g. Scheuer \cite{scheuer74}, Falle \cite{falle91}) which expands first in the
Interstellar Medium (ISM) and later in the Intergalactic Medium
(IGM) of the host galaxy. 
The evolution of such cocoons is traced by the radio lobes, which in
themselves are only a (albeit important) 'side effect' caused by the
presence of magnetic fields in the cocoon.  

The giant radio galaxies (GRGs) are those radio sources whose lobes span a
(projected) distance of above 1 Mpc\footnote{We use $H_0 =
50$~km\,s$^{-1}$\,Mpc$^{-1}$ and $q_0 = 0.5$ throughout this paper.}. 
As such, GRGs must represent a late phase in the evolution of radio
sources. 
Models of radio source evolution (e.g. Kaiser et al. \cite{kaiser97} and
Blundell et al. \cite{blundell99}) predict the radio power and linear
size evolution of powerful radio sources with time. 
According to these models, GRGs must be extremely old (i.e. typically
older than $10^8$ yr) and probably also located in underdense
environments, as compared to smaller radio sources of comparable radio
power (e.g. Kaiser \& Alexander \cite{kaiser99}). 

Multi-frequency radio observations (Mack et al. \cite{mack98}) have shown that
spectral ages of GRGs are of the same order as expected from source
evolution models. It is, however, not clear at all whether {\sl spectral}
ages are representative of the {\sl dynamical} ages (e.g. Parma et
al. \cite{parma99}). This questions the validity of radio-based
determinations of the 
properties of the environments of these sources. Still, constraints on
the environments of GRGs are of high importance since the radio lobes of
these sources penetrate deeply into the intergalactic medium. 
It is almost impossible to find the
properties of this medium, otherwise than from studies of such radio lobes
(e.g. Subrahmanyan \& Saripalli \cite{subrahmanyan93}). 

A major problem for such studies is that currently known GRGs have not
been uniformly selected. The difficulties encountered while selecting
extended radio sources has been demonstrated by Saunders et al. 
(\cite{saunders87}), who searched for GRGs in a small
region of the 151-MHz 6C survey. The 6C survey, with only 30~mJy
beam$^{-1}$ {\sc rms}-noise and a beamsize of
$4.2\arcmin\times4.2\arcmin\,{\rm cosec}\,\delta$ FWHM (with $\delta$ the
declination) has an excellent sensitivity to large, faint objects. 
However, using higher resolution observations they found that only at
integrated flux densities above 5~Jy has a radio source larger than
$5\arcmin$ a good chance of being a genuine GRG in the 6C. At a flux
density level of 1 Jy, Saunders et al. find that most of the sources which
appear as large extended structures on the 6C survey maps are the result of
confusion of physically unrelated sources. 
Their work demonstrates that 1) an efficient 
search for GRGs has to be done with sufficient angular
resolution to minimize confusion problems and that 2) it should
be done with a high sensitivity, also for large-scale structures (up to
a few tens of arcmin) on the sky. The recently completed WENSS survey
(Rengelink et al. \cite{rengelink97}) meets both these demands.

In this paper we report of the selection of new giant radio sources from
the WENSS.  Subsequent papers will present additional radio observations
(Schoenmakers et al. \cite{schoenmakers00b}), a more detailed 
analysis of the spectroscopic data and a discussion of the evolution
of GRGs (in preparation, but see Schoenmakers \cite{schoenmakers99a}).  
In Sect. \ref{sec2:method} we outline the selection technique and
criteria. Section \ref{sec2:identification} presents the strategy we have
adopted for finding the optical identifications, and 
Sect. \ref{sec2:spectroscopy} describes the spectroscopic
observations of these identifications. In Sect. \ref{sec2:first-results}
we present the first results of the new sample of GRGs: Flux
densities, linear sizes, redshifts, etc. A discussion of these results
and on the sensitivity of the WENSS survey to extended radio sources is
given in Sect. \ref{sec2:discussion}.    

Throughout this paper, a spectral index $\alpha$ is defined according
to the relation $S_{\nu} \propto \nu^{\alpha}$ between flux density
$S_{\nu}$ at frequency $\nu$, and the frequency $\nu$.

\begin{table*}[t!]
\centering
\caption{\label{DS1923:known_grgs} List of all previously 
known GRGs (i.e. before 1998) in
the area of the sky covered by the WENSS and fulfilling our selection
criteria. Column 1 gives the name of the source in IAU-format; column
2 gives the more common name of the source, if available; column 3 and
4 give the coordinates of the host galaxy in B1950.0; column 5 and 6
give the redshift of the host galaxy and a reference to it; column 7
gives the projected linear size of the radio source in Mpc; column 8
designate the morphological type of the radio source, i.e. FRI or FRII
or an intermediate type, FRI/II. If a `B' is added, it means that the
host galaxy is a broad-line object, a `Q' means that it is a
quasar. Column 9 gives a reference to radio maps in which the radio
morphology can be studied. Columns 10 and 11 give the flux density at
325 MHz and the reference for this value.}
\setlength{\tabcolsep}{5.8pt}
\begin{tabular}{l l l l l l c l l r@{$\,\pm\,$}r l} 
\hline\hline\\
\multicolumn{1}{c}{(1)} & \multicolumn{1}{c}{(2)} &
\multicolumn{1}{c}{(3)} & \multicolumn{1}{c}{(4)} &
\multicolumn{1}{c}{(5)} & \multicolumn{1}{c}{(6)} &
\multicolumn{1}{c}{(7)} & \multicolumn{1}{c}{(8)} &
\multicolumn{1}{c}{(9)} & \multicolumn{2}{c}{(10)} &
\multicolumn{1}{c}{(11)} \\
IAU Name & Other name & \multicolumn{1}{c}{R.A.} &
\multicolumn{1}{c}{Dec.} &  \multicolumn{1}{c}{$z$} & \multicolumn{1}{c}{Refs.} & \multicolumn{1}{c}{Size} & \multicolumn{1}{c}{Type} & \multicolumn{1}{c}{Refs.} &
\multicolumn{2}{c}{$S_{325}$} & \multicolumn{1}{c}{Refs.} \\
\multicolumn{2}{c}{\ } & \multicolumn{2}{c}{(B1950.0)} & & &
\multicolumn{1}{c}{$[$\,Mpc\,$]$} & FR & & \multicolumn{2}{c}{$[\,$Jy$\,]$} &
\\
\hline \\[1ex]
\object{B\,0050$+$402} & & 00 50 45.1 & 40 11 10.0 & 0.1488 & DJ95 & 1.5 &
II & V89 & 1.41 & 0.04 & WE \\ 
\object{B\,0055$+$300} & NGC\,315 & 00 55 05.6 & 30 04 56.8 & 0.0167 & C75
& 1.7 & I/II & W81 & 9.71 & 0.18 & M97\\
\object{B\,0104$+$321} & 3C\,31   & 01 04 39.2 & 32 08 44.0 & 0.0169 & S97
& 1.3 & I & S83 & 13.67 & 0.28 & WE \\
\object{B\,0109$+$492} & 3C\,35 & 01 09 04.1 & 49 12 40.1 & 0.06701 & B72
& 1.1 & II & B82 & 7.19 & 0.15 & WE \\
\object{B\,0136$+$396} & 4C\,39.04 & 01 36 33.6 & 39 41 51.2 & 0.2107 &
S73 & 1.6 & II & FB78,H79 & 4.75 & 0.10 & WE \\
\object{B\,0157$+$405} & 4C\,40.09 & 01 57 22.4 & 40 34 34.2 & 0.0827 &
DJ95 & 1.9 & I/II & V89 & 2.98 & 0.08 & WE \\  
\object{B\,0309$+$411} & & 03 09 44.8 & 41 08 48.7 & 0.134 & B89 & 1.8 &
II-B & B89 & 1.38 & 0.04 & WE \\
\object{B\,0745$+$560} & DA\,240 & 07 45 46.1 & 56 01 56.2 & 0.0356 & W74
& 2.0 & II & S81 & 17.05 & 0.35 & M97 \\
\object{B\,0821$+$695} &         & 08 21 01.9 & 69 30 25.8 & 0.538  & L93
& 3.0 & II & L93 & 0.63 & 0.02 & WE \\
\object{B\,0945$+$734} & 4C\,73.08 & 09 45 09.9 & 73 28 22.2 & 0.0581 &
D70 & 1.5 & II & J86 & 10.43 & 0.21 & WE \\
\object{B\,1003$+$351} & 3C\,236 & 10 03 05.4 & 35 08 48.0 & 0.0989 & S85
& 5.7 & II & S80  & 13.13 & 0.26 & M97 \\
\object{B\,1029$+$570} & HB\,13 & 10 29 48.2 & 57 00 45.8 & 0.045 & S96 &
2.6 & I & MA79 & 1.09 & 0.04 & WE \\
\object{B\,1209$+$745} & 4CT\,74.17 & 12 09 36.0 & 74 35 45.5 & 0.107 & M79
& 1.2 & II & B81 & 2.06 & 0.05 & WE \\
\object{B\,1309$+$412} & & 13 09 27.1 & 41 14 53.5 & 0.1103 & D90 & 1.0 &
II & V89 & 1.61 & 0.04 & WE \\
\object{B\,1312$+$698} & DA\,340 & 13 12 22.1 & 69 53 10.0 & 0.1060 & S87
& 1.3 & II & S87 & 4.16 & 0.09 & WE \\
\object{B\,1358$+$305} & & 13 58 29.3 & 30 33 47.7 & 0.206  & P96 & 2.6 &
II & P96 & 1.84 & 0.04 & P96 \\
\object{B\,1626$+$518} & & 16 26 48.5 & 51 53 05.0 & 0.0547 & G92 & 1.6 & II-B
& R96 & 1.62 & 0.03 & WE \\
\object{B\,1637$+$826} & NGC\,6251 & 16 37 57.0 & 82 38 18.6 & 0.023 & W77
& 3.0 & I/II & P84 & 11.55 & 0.23 & M97 \\
\object{B\,2043$+$749} & 4C\,74.26 & 20 43 13.0 & 74 57 08.7 & 0.104 & R88 & 1.6 & II-Q
& R88 & 4.76 & 0.10 & WE \\
\hline \hline \\
\end{tabular}
\begin{minipage}{\linewidth}
References: B72: Burbidge \& Strittmattar \cite{burbidge72};
B81: van Breugel \& Willis \cite{breugel81}; 
B82: van Breugel \& J\"{a}gers \cite{breugel82};
B89: de Bruyn \cite{debruyn89};
C75: Colla et al. \cite{colla75};
D70: Demoulin \cite{demoulin70};
D90: Djorgovski et al. \cite{djorgovski90};
DJ95: Djorgovski et al. \cite{djorgovski95};
FB78: Fomalont \& Bridle \cite{fomalont78};
G92: de Grijp et al. \cite{degrijp92};
H79: Hine \cite{hine79};
J86: J\"{a}gers \cite{jagers86};
M79: Miley \& Osterbrock \cite{miley79};
MA79: Masson \cite{masson79};
M96: Marcha et al. \cite{marcha96};
M97: Mack et al. \cite{mack97};
P84: Perley, Bridle \& Willis \cite{perley84};
P96: Parma et al. \cite{parma96};
R88: Riley et al. \cite{riley88};
R96: R\"{o}ttgering et al. \cite{rottgering96};
S73: Sargent \cite{sargent73};
S80: Strom \& Willis \cite{strom80};
S81: Strom et al. \cite{strom81};
S82: Saunders \cite{saunders82};
S83: Strom et al. \cite{strom83};
S85: Spinrad et al. \cite{spinrad85};
S86: Saripalli et al. \cite{saripalli86};
S87: Saunders et al. \cite{saunders87};
S97: Simien \& Prugniel \cite{simien97};
V89: Vigotti et al. \cite{vigotti89};
WE : WENSS (measured in the radio map);
W74: Willis et al. \cite{willis74};
W77: Wagett et al. \cite{wagett77};
W81: Willis et al. \cite{willis81}.
\end{minipage}
\end{table*}

\section{Sample selection}
\label{sec2:method}

\subsection{The WENSS survey}

The Westerbork Northern Sky Survey is a 325-MHz survey of the sky above
$+28\degr$ declination. About a quarter of this area has also
been observed at a frequency of 609 MHz. The unique aspect of WENSS is that
it is sensitive to spatial structures over 1 degree on the sky at 325 MHz. The limiting flux density to
unresolved sources is about 15 mJy ($5\sigma$) and the FWHM of the
beam is $54\arcsec\times 54\arcsec {\rm cosec}\,\delta$,
with $\delta$ the declination.   A detailed description of the observing
and data reduction techniques used can be found in Rengelink et
al. (\cite{rengelink97}).  
The sky area above $+74^{\circ}$ declination has been observed 
with an increased total bandwidth, so that the
limiting flux density to unresolved sources is about 10 mJy
($5\sigma$) in this sky area. 

Using the Wieringa (\cite{wieringa91}) source counts at 325-MHz, the flux density at
which the same amount of confusion in the WENSS as Saunders et al. (\cite{saunders87})
encountered in the 6C survey can be calculated. For a typical radio source
spectral index of $-0.8$, a similar amount of confusion can be expected in
the WENSS at a flux density of $\sim\!400$ mJy, which is almost seven
times lower than that for the 6C survey.  Similarly, it can be shown that
confusion would dominate the selection sources only below 20 mJy in the
WENSS survey, which is below its
completeness limit of $\sim 30$ mJy (Rengelink et al. \cite{rengelink97}).  
This implies that we should be able to efficiently find GRGs in the WENSS
down to relatively low flux density levels. 

\subsection{Selection criteria}

In order for a source to be a candidate low redshift GRG, we have used
the following criteria. A candidate GRG must have:
\begin{enumerate}
\item an angular size larger than 5 arcminute, and 
\item a distance to the galactic plane of more than 12.5 degree.
\end{enumerate}
The angular size lower limit of $5\arcmin$ is the size at which some
basic morphlogical information of a source can still be obtained at all
declinations the survey has covered. this  
corresponds to a physical size of $\sim\!750$~kpc at $z=0.1$,
$\sim\!1300$~kpc at $z=0.2$ and $\sim\!1700$~kpc at $z=0.3$, and will
therefore introduce a redshift-dependent linear size bias in the
sample.
To avoid high galactic extinction values and confusion by a large surface
density of foreground stars, we have restricted ourselves to galactic
latitudes above $12.5\degr$. This results in a survey area of
$\sim\!2.458$ steradian ($\sim\!8100^{\sq}$).

Tab. \ref{DS1923:known_grgs} presents all previously discovered GRGs
whose angular size and position on the sky agree with the above selection
criteria. The majority of these are smaller than 2~Mpc in size. If their sizes
are characteristic for the whole population of GRGs, we
thus expect that the majority of selected sources will have a redshift below
$\sim\!0.35$. Assuming that the host galaxies are not less luminous than
those of the LRL sample of powerful radio sources (Laing et
al. \cite{laing83}) they should be identifiable on the Digitized POSS-I
survey (DSS).  

\subsection{Selection method}

Candidate radio sources were selected using a visual inspection of the WENSS radio maps.
We preferred this method over possibly more objective, machine
controlled selection methods because the complexity of the WENSS radio
maps (i.e the high source surface density and the unavoidable presence of
spurious artefacts such as low-level sidelobes of bright sources, etc.)
and the wide variety in possible morphologies would make it very difficult
to tune such an algorithm. Looking at the maps allows one to easily
recognize low-level extended structures in a crowded field.  

Above declination $+74\degr$ we have initially selected our candidates
using the earlier available NVSS survey maps (Condon et
al. \cite{condon98}), but we subsequently repeated
the selection using the WENSS maps.  We found that no WENSS selected
candidates were omitted using the NVSS.  On the contrary, we have
identified two NVSS sources (B\,1044$+$745 and B\,0935$+$743) that we most
likely would not have selected from the WENSS survey alone due to their
faintness in the latter. We will elaborate on this when we discuss the
selection effects (Sect. \ref{sec2:sensitivity-limit}).

\subsection{Removing confused sources}

The declination dependent beam size results in an
unavoidable increase of confusion with decreasing declination. Only with higher angular resolution observations can we determine whether such sources are separate unrelated radio sources. We have used the following additional sources of radio data to achieve this:

First, where available, we have used the 612-MHz WENSS maps which have
twice the resolution of the 325-MHz maps. Also the 1.4-GHz NVSS survey, which
does not have a declination-dependent beam size and which covers almost
the entire area of the WENSS survey, is highly usefull in this respect. 
Furthermore, we have used the much higher resolution ($5\farcs4$ FWHM)
maps from the 1.4-GHz FIRST survey (Becker et al. \cite{becker95}) where
available. Also the FIRST survey has 
mapped a large fraction of the area observed by WENSS, notably the lower
declination range away from the galactic plane. 
Finally, for candidates in areas of the sky where the FIRST survey was
not (yet) available, we obtained short 1.4-GHz WSRT observations. 

A consequence of the different methods used to eliminate confused
sources is that the angular size of the objects are not well determined in
all cases. There are two important factors which influence such estimates:
First, for edge-brightened sources (FRII-type) high-resolution observations
would be required for an accurate measurement, but we do not have these
for all such sources and even the ones we have differ in quality and
resolution. Second, for FR-I type sources the angular size measured on a
map depends strongly on surface-brightness sensitivity. Therefore, sources
may have been accidentally removed from the sample
because of wrongly estimated sizes.

\subsection{Identification of the host galaxies} 
\label{sec2:identification}

We have used the digitized POSS-I survey (the `Digitized Sky Survey',
DSS) and, in a later stage, also the digitized POSS-II survey to
identify the host galaxies of the selected radio sources.  The
magnitude limit of the red POSS-I plates is $\sim\!20-20.5$; the
POSS-II is somewhat more sensitive.

Adopting the Cousin R-band magnitude -- redshift relation for the host
galaxies of the radio sources in the LRL sample (Dingley \cite{dingley90}), we
expect to be able to identify host galaxies out to $z\sim\!0.5$ (note that the
transmission curves of the Cousin R-band and the POSS-E band are much
alike). 

To identify the host galaxy of a radio source which extends over several
arcminutes, a radio core position is often necessary.  We have
used the WENSS, NVSS, FIRST and our own WSRT radio observations to
identify radio cores of the selected sources.  For many sources we indeed
find a compact central radio source coincident with an optical
galaxy in the POSS-I and/or POSS-II. 

We cannot rule out that for an individual source the so-found optical galaxy 
is an unrelated foreground galaxy, and that the actual host galaxy of the
radio source is a much farther and fainter galaxy. However, for the sample
as a whole, we believe this to be only a minor problem: It would make the
radio sources even larger than they already are and the chance of such an
occurrence is very small anyway. 

For the source B\,1918$+$516 the POSS-I plates did not show an obvious
host galaxy candidate. Therefore, an optical CCD image has been made by
P.N. Best using the LDSS imaging spectrograph on the 4.2-m WHT telescope
on La Palma.  This image (see Fig. \ref{DS1923:1918+516}) reveals a faint
galaxy, close to a relatively bright star. We believe this galaxy to be
the host galaxy, due to its proximity to the radio core. 

We cross-correlated the positions of the optical galaxies with the NASA
Extragalactic Database (NED). In case an optical source with known redshift is
found and the resulting size of the radio source is below 1 Mpc, we
removed that source from our sample.
We present the list of remaining (i.e. after removing sources identified
as non-GRGs on basis of NED data) candidate GRG sources in
Tab. \ref{tab:big_list_of_candis}. We provide IAU-formatted source names,
approximate coordinates of the radio sources, WENSS flux densities,
approximate angular sizes and whether we are convinced this a genuine
giant radio galaxy candidate on basis of its radio morphology. 
We remark that many of the WENSS selected sources were rejected  
after a look at the maps of these sources from the NVSS survey.  

\section{Optical spectroscopy}
\label{sec2:spectroscopy}

Optical spectra of a large fraction of possible host galaxy of the
candidates have been obtained. In order to construct an as complete as
possible flux-density limited sample from our candidates, we have tried to
obtain spectra and redshifts for all candidate sources 
in Tab. \ref{tab:big_list_of_candis} with a 325-MHz flux density above 1
Jy \footnote{
From a recent paper by Lara et al. (\cite{lara01}), however, it was noted that
three genuine candidate sources 
were missed after a first selection round from WENSS (B\,1838+658,
B\,1919+741, B\,0603+612). The source B\,1838+658 has a
redshift of $0.23$ (NED) and is as such a GRG; for B\,1919+741 the
redshift is $0.194$ and for B\,0603+612 it is $0.227$ (Lara et
al. \cite{lara01}). 
Also, the source B\,1855+310 was not further observed.
Further, high resolution observations by Lara et al. show that some
sources, whose size was estimated to be less than $5\arcmin$ on basis of
the WENSS maps, actually have angular sizes $\gtrsim 5\arcmin$ and should
thus be included in our candidate sample.  
The implication of these findings is that the earlier made claim  that we
have observed a complete flux-density limited sample of GRGs (paper
II, Schoenmakers et al. \cite{schoenmakers00b}), must be recalled.}. 
 
In Tab. \ref{DS1923:INT_observations} we present the log of the
spectroscopic observations.  
On the 2.5-m INT telescope on
La Palma we used the IDS-235 camera with the Ag-Red collimator and the R300V
grating. The camera was equipped with a 1k$\times$1k TEK chip. This
setup results in a total wavelength coverage of $\sim\!3500$\AA~and a
pixel scale of $\sim\!3.2$\AA/pixel in the dispersion direction and
$0\farcs74$/pixel in the spatial direction. 
Depending on the magnitude estimated redshift of the host galaxy candidate, 
the central wavelength of the spectrograph was
set at either 5500, 6000 or 6500\AA.  The
slit-width was held constant at $2\arcsec$. Only in the case of the source
B\,0809$+$454 we used a $3\arcsec$-wide slit because of uncertainties
in the optical position at the time of observation.  Flat-field
observations were made at the beginning of each night by using the
internal Tungsten lamp. Wavelength calibration was achieved by
internal arc-lamp exposures (from {\sc cu-ar} and {\sc cu-ne} lamps)
which were taken at the beginning of each night and immediately
after each object exposure.  For absolute flux calibration and to
correct for the wavelength-dependent sensitivity of the instrument,
several spectroscopical standard stars were observed each night, using
a $5\arcsec$-wide slit.

The spectra have been reduced using the NOAO {\sc iraf} data reduction
software.  One dimensional spectra have been extracted using a
$4\arcsec$-wide aperture centered on the peak of the spatial profile
of the identification. The resulting spectra are shown in
Figs. \ref{DS1923:0211+326}-\ref{DS1923:2147+816}.
For each object exposure the wavelength
calibration has been checked against several bright sky-lines in a non
background-subtracted frame. 

In addition to these observations, a spectrum of the source 
B\,1450$+$333 was obtained on 8 July 1997 by P.N. Best, using the 4.2-m WHT
 on La Palma, equipped with the ISIS spectrograph. An
optical CCD image and a spectrum of the source B\,1543$+$845 were obtained
by I.M. Gioia on 4 and 5 March 1998 using the HARIS imaging spectrograph
on the 2.2-m University of Hawaii Telescope on Mauna Kea, Hawaii.

Table \ref{DS1923:redshifts} presents the measured redshifts and the
spectral features used to determine these. The optical images and spectra
of the observed host galaxies are presented in Figs. 
\ref{DS1923:0211+326}--\ref{DS1923:2147+816}. For the source B\,1245+676 a
high-quality spectrum and the therefrom derived redshift had been
published earlier by Marcha et al. (\cite{marcha96}). We have therefore not observed
the host galaxy of this radio source again. We have, however, reobserved the
source B\,1310+451 in order to obtain a higher quality spectrum of the
host galaxy of this source.

Several other radio and optical properties of the observed candidates, and
of the three sources that were found to be GRGs from available data in the
NED (B\,1144+352, B\,1245+676 and B\,1310+451) are given in
Tab. \ref{DS1923:radiocores} and \ref{DS1923:radioproperties}.

\section{First results}
\label{sec2:first-results}

\subsection{The number of GRGs}

Of the 33 candidate sources which we have been able to identify
spectroscopically, only three,
and possibly five have projected linear sizes below 1 Mpc. These are
B\,0217$+$367, B\,1709$+$465 and B\,1911$+$479. The uncertain cases
are the source B\,0905$+$352, for which we do not have a well
determined redshift yet and B\,1736$+$375 which may consist of two
unrelated radio sources.  
Together with the 19 known GRGs in the area of the
WENSS which share our selection criteria, we have so far identified 47 GRGs.
This is by far the largest sample of sources with a projected
linear size above 1 Mpc selected from a single survey. 
We also mention the existence of the large, but highly incomplete sample of GRGs which has been compiled by Ishwara-Chandra \& Saikia (\cite{ishwara-chandra99}) from the literature. 
Although of similar size, the sample presented in this paper is
better suited for statistical investigations of the
radio and optical properties of GRGs since it has been selected from a
single survey and in a more uniform matter.

\subsection{Distribution functions of the sample}

\subsubsection{Flux density distribution}

\begin{figure*}[t!]
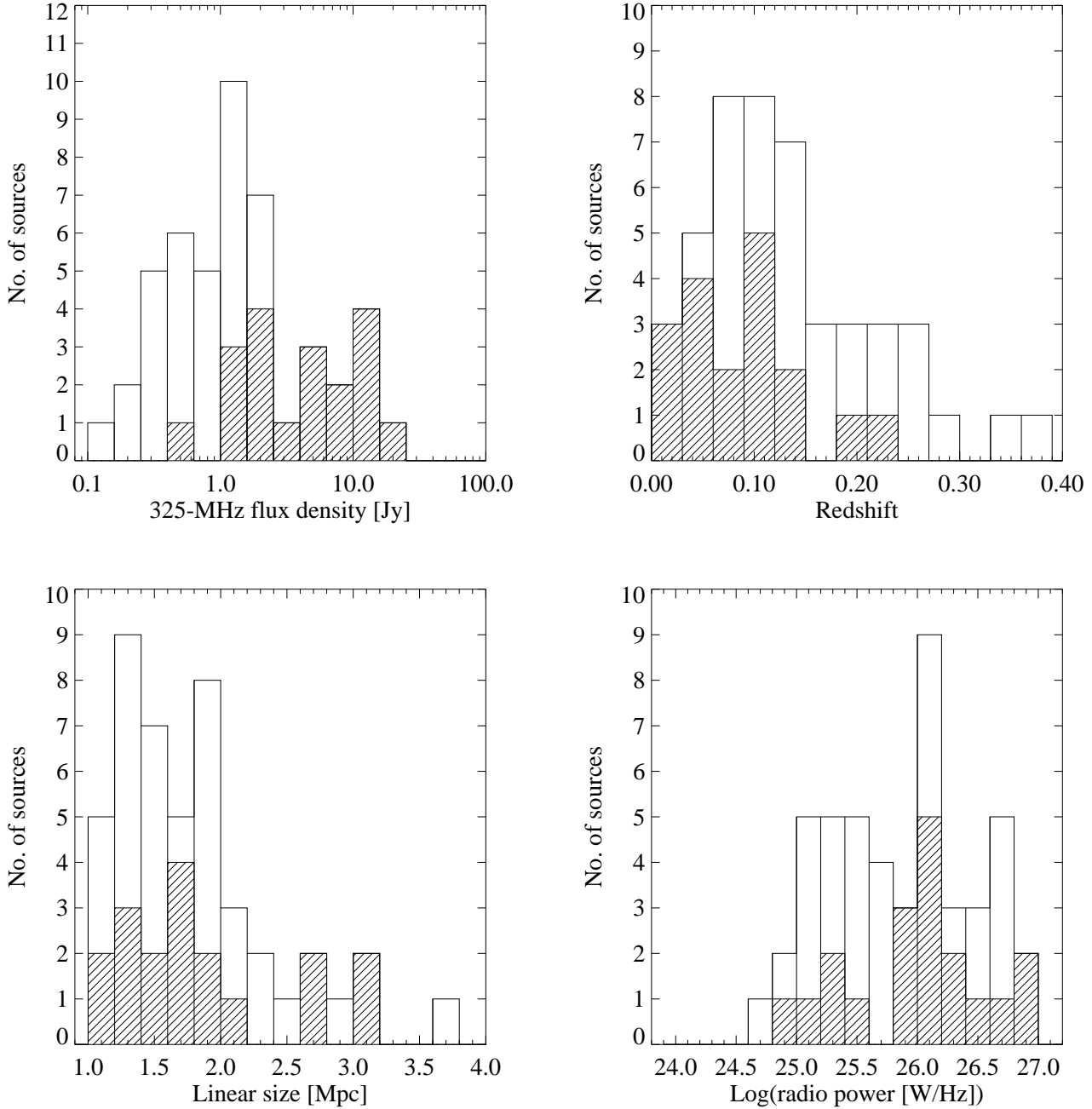

\setlength{\tabcolsep}{0pt}
\begin{tabular}{l l}
\resizebox{0.5\hsize}{!}{\epsfig{file=DS1923.1a,angle=90}} &
\resizebox{0.5\hsize}{!}{\epsfig{file=DS1923.1b,angle=90}} \\
\resizebox{0.5\hsize}{!}{\epsfig{file=DS1923.1c,angle=90}} &
\resizebox{0.5\hsize}{!}{\epsfig{file=DS1923.1d,angle=90}} \\
\end{tabular}
\caption{\label{DS1923:distributions} Histograms of several properties of
the new and old sample of GRGs. In all plots, the hatched bars
indicate the distribution of previously known GRGs (see
Tab. \ref{DS1923:known_grgs}). {\bf a} (upper left): The 325-MHz flux
density distribution of the GRGs. We have used a binsize of $0.2$ in the
logarithm of the flux density in Jy.  {\bf b} (upper right): The
redshift distribution of the GRGs, using a binsize of $0.03$ in redshift. The
source 8C\,0821$+$695 at $z=0.538$ lies outside the range of this plot.  
{\bf c} (lower left): The projected linear size distribution of the GRGs,
using a binsize of 0.2 Mpc. The source 3C\,236 ($D=5.7$ Mpc) lies
outside the range of this plot.  
{\bf d} (lower right): The rest-frame 325-MHz
radio power distribution of the GRGs, using bins of width 0.2 in the
logarithm of the radio power in W\,Hz$^{-1}$.}
\end{figure*}

Fig. \ref{DS1923:distributions}a presents a
histogram of the 325-MHz flux density distribution of the sample
of 47 sources. The hatched bars indicate the flux density distribution of the sample of previously known GRGs. 
Not surprisingly, we find that all newly discovered GRGs have flux
densities below 3 Jy, only; all ten GRGs with a 325-MHz flux density
above this value had already been identified as such. 
For the area of the sky covered by WENSS, we can therefore agree with Riley (\cite{riley89}) who argued that no bright extended sources are missing from the LRL sample of bright radio sources ($S_{178}\!>\!10.9$ Jy, Dec.\,$>\!+10\degr$, $|b|\!>\!10\degr$). 
Furthermore, we have extended the range of flux densities at which low
redshift GRGs are found to sub-Jy levels. The median flux
density of the combined sample is 1.15 Jy, which is almost a factor of four
below the median value of the sample of previously known GRGs (see
Tab. \ref{DS1923:medians}). 

\subsubsection{Redshift distribution}

The new sample of GRGs mostly contains GRGs at $z \la 0.3$ (see
Fig. \ref{DS1923:distributions}b). The only exceptions are the sources
8C\,0821$+$695 ($z=0.538$; Lacy et al. \cite{lacy93}), B\,0750$+$434
($z=0.347$) and B\,0925$+$420 ($z=0.365$). The redshift distribution peaks
at $z \sim 0.1$\,.  The decrease in the number of sources towards higher 
redshift is likely to be due to the lower angular size limit used in
the selection of the sample. Further, only more powerful GRGs will be
selected towards higher redshift, of which the space density is likely to
be lower. The median redshift of the sample 
of new GRGs is 0.1404, which is higher than that of the old GRGs alone
(0.099; see Tab. \ref{DS1923:medians}). Since the average flux density is
lower this is not surprising. The median redshift of the combined 
sample is 0.1175\,. 

\subsubsection{Linear size distribution}
 
No source with a (projected) linear size exceeding that of the GRG 3C\,236 has been found. This source is therefore still the largest known radio galaxy in the Universe.
In Fig. \ref{DS1923:distributions}c we have plotted the (projected)
linear size distribution of our sample. 
The median values of the
linear sizes of the old and new GRGs lie close together (see
Tab. \ref{DS1923:medians}).  The majority of sources have linear sizes
between 1 and 2 Mpc; the distribution of the combined
sample falls off strongly at a linear size of 2 Mpc. This sharp
decrease was not clear from the sample of `old' GRGs only, due to the
small number of sources.
In Sect. \ref{sec2:cutoff} we will discuss whether the observed cut-off is
a result of selection effects or an intrinsic property of the population
of GRGs. 

\subsubsection{Radio power distribution}

We calculated the emitted radio power at a rest-frame frequency
of 325 MHz assuming isotropic emission. 
Since the redshifts are low, the radio K-correction is only small. 
For the new GRGs the measured
spectral index between 325 and 1400 MHz has been used; for the `old' GRGs a spectral index of $-0.8$ has been assumed if no reliable literature value could be found. The distribution of 325-MHz radio powers has been plotted in
Fig. \ref{DS1923:distributions}d. Despite the fact that a
large number of GRGs has been found at flux density values well below that of the `old' sample, the distribution of the radio
powers of the new sources largely overlaps that of the old sources.  
This is related to the, on average, higher redshifts of the new sources.
The combined sample is distributed rather uniformly between $10^{25}$
and $10^{27}$ W\,Hz$^{-1}$; the sharp peak at the radio power bin
centered on
$10^{26.1}$ W\,Hz$^{-1}$ is most likely a result of small number
statistics. At 325 MHz, the traditional separation between FRII and FRI-type
sources (Fanaroff \& Riley \cite{fanaroff74}) lies near $10^{26}$ W\,Hz$^{-1}$,
which is close to the median value of the combined sample (see
Tab. \ref{DS1923:medians}).

\begin{table}
\caption{\label{DS1923:medians} Median values of some of the properties of
the GRGs in the sample of `old' GRGs, the sample of new GRGs and the
combined sample of 47 sources. Column 1 gives the property; column 2
to 4 give the median value of these properties for the old, new and
combined sample, respectively.}  
\begin{tabular}{l l l l}
\hline\hline\\
\multicolumn{1}{c}{(1)} & \multicolumn{1}{c}{(2)} & \multicolumn{1}{c}{(3)} & \multicolumn{1}{c}{(4)}\\
\multicolumn{1}{c}{Property} & \multicolumn{3}{c}{\hrulefill~Median values~\hrulefill} \\
 &  \multicolumn{1}{c}{Old} & \multicolumn{1}{c}{New} &  \multicolumn{1}{c}{Comb.} \\
\hline \\
$S_{325}~[$\,Jy\,$]$ & 4.160  & 0.685  & 1.150  \\ 
$\log(P_{325}~[$\,W Hz$^{-1}\,])$ & 26.041 & 25.625 & 25.951 \\
Size $[$\,Mpc\,$]$   & 1.65  & 1.63  & 1.64  \\
Redshift & 0.099 & 0.1404 & 0.1175 \\
\hline \\
\end{tabular}
\end{table}

\section{Discussion}
\label{sec2:discussion}

\subsection{The `sensitivity limit' of WENSS}
\label{sec2:sensitivity-limit}

\begin{figure}
\resizebox{\hsize}{!}{\epsfig{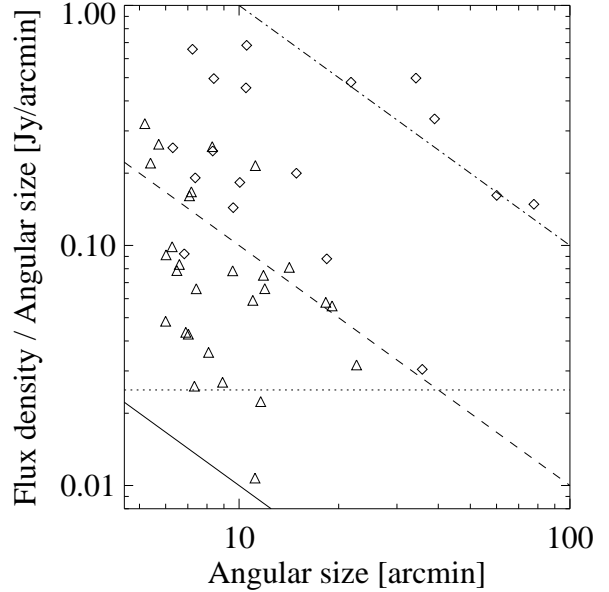}}
\caption{\label{DS1923:signal-to-noise}
A plot of the 325-MHz flux density divided by the angular size against the
angular size for the `old' (diamonds) and the newly discovered GRGs
(triangles). The diagonal lines indicate a constant integrated flux
density and are drawn for 0.1 (solid), 1 (dashed) and 10 (dot-dashed)
Jy. From this plot we determine the sensitivity limit of our selection method, 
($S_{int}/\theta_{max} = 0.025$ Jy/arcmin), indicated by the dotted
horizontal line.}
\end{figure}

For a radio source to be included in our sample, it must be larger than
$5\arcmin$ and it must have been noticed on the radio maps as a large radio
source. The latter is related to a surface brightness
criterion: the average
surface brightness, or integrated signal-to-noise ratio, must be high
enough to be detected as a single radio source structure. The integrated signal-to-noise ratio, $(S/N)_{int}$, for a resolved radio source is given by
\begin{equation}
\left(\frac{S}{N}\right)_{int} \approx \frac{S_{int}}{\sigma \sqrt{A}},
\label{eq2:s/n_1}
\end{equation}
where $S_{int}$ is the integrated flux density of the source and the
surface area $A$ is in units of beams. The surface area can be
rewritten as $A = c \cdot \theta_{max}^2$, where $\theta_{max}$ is the
angular size (major axis) of the radio source and $c$ is a number that
relates the angular size to the surface area (cf. the length-to-width
ratio).  For instance, for an elliptically shaped radio source with a
length-to-width ratio of 3, $c = \frac{\pi}{(12 \ln 2)} \cdot
\theta_{beam}^2$ with $\theta_{beam}$ the (FWHM) beam-size of the
observation.  If we substitute the above expression for $A$ in
Eq. \ref{eq2:s/n_1} we find
\begin{equation}
\left(\frac{S}{N}\right)_{int} \propto \frac{S_{int}}{\theta_{max}}~.
\end{equation}
Fig. \ref{DS1923:signal-to-noise} shows $S_{int}/\theta_{max}$
against $\theta_{max}$ for the sources in our sample which we have
identified as GRGs. We find that the lowest values of
$S_{int}/\theta_{max}$ for selected sources lie in the range between
$0.02 - 0.03$ Jy/arcmin; the source which lies well below this
line is B\,1044$+$745, which is one of the two sources that were selected
only for its radio structure in the NVSS and should therefore be
situated below the WENSS `sensitivity' limit.  The other source which lies
just below the line is B\,1245$+$676, which was selected from the
WENSS but can be considered a `border-line' case. The two sources just
above the limit are B\,0935$+$743 and B\,1306$+$621.

The sensitivity limit appears to be almost independent of angular
size at least up to a size of $\sim\!40$ arcminute.
The sensitivity of WENSS to objects with an
angular size above 1 degree on the sky decreases, so that the sensitivity
limit inevitably must rise eventually. 

From the figure we conclude that sources with
$\theta_{max} \ge 5\arcmin$ will most likely be selected
if $S_{int}/\theta_{max} \ga 0.025$ Jy/arcmin.  We will use this
criterium to specify the regions in the radio power -- linear size --
redshift $(P,D,z)$ parameter space which is accessible by our selection. 
 
\subsection{The radio power -- linear size diagram}

\begin{figure}
\resizebox{\hsize}{!}{\epsfig{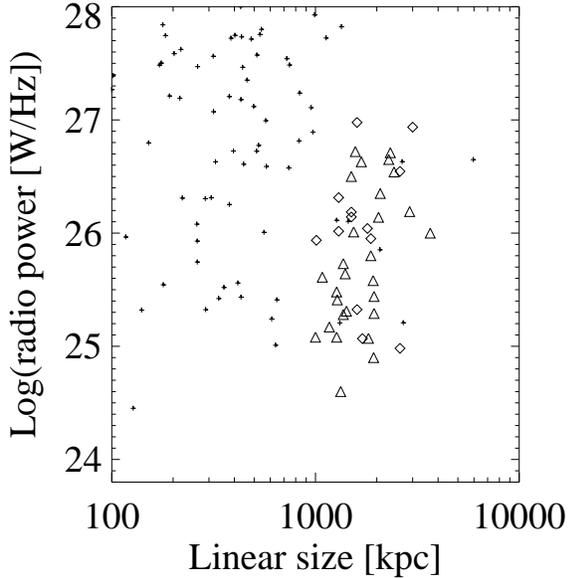}}
\caption{\label{DS1923:p-d_all-diagram}
The 325-MHz radio power against linear size ($P-D$ diagram) for
the formerly known GRGs not part of the LRL sample (diamonds), the
newly discovered GRGs (triangles) and sources from the LRL sample with
$z<0.6$ (plusses).} 
\end{figure}

In Fig. \ref{DS1923:p-d_all-diagram} we have plotted for all identified GRGs in our sample the linear size,
$D$, against the radio power at 325 MHz, $P$, the so-called $P-D$
diagram.  For reference, all
sources of the LRL sample with $z < 0.6$, which is the same redshift
range as in which the GRGs are found, are plotted as well. 
Note that several of the
formerly known GRGs are part of the LRL sample; these have been
plotted as LRL sources.

From this plot the following can be concluded.  First, although we
have conducted the most extensive systematic search for GRGs to
date, there are no sources in the upper right part of
the $P-D$ diagram, i.e. the region occupied by sources with large size
and high radio power. If such sources had existed in our search area,
They would most likely have discovered because of their inevitable high flux
density. Second, the few GRGs which have a linear size above 2 Mpc have, on average, a higher radio power than smaller-sized GRGs.

\begin{figure}
\hspace{-0.5cm}\resizebox{\hsize}{!}{\epsfig{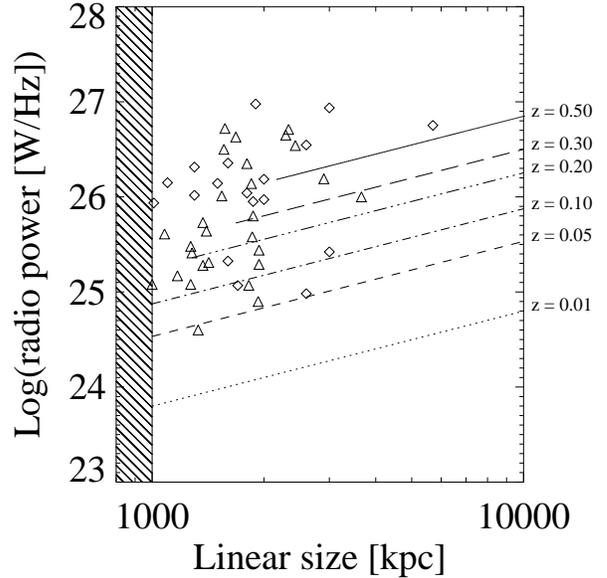}}
\caption{\label{DS1923:p-d-selected} $P-D$ diagram, filled with formerly known GRGs
(diamonds) and newly discovered GRGs (triangles). The lines
indicate the lower radio power limit for a source at a particular
redshift as a function of linear size; the area directly above such a
line is the accessible region in the $(P,D)$ parameter space at that
redshift. See text for further details.}
\end{figure}

To investigate which region of the $P-D$ diagram is accessible
through our WENSS selection, we have plotted in
Fig. \ref{DS1923:p-d-selected} lines which represent the lower sensitivity
limit at constant redshift.  Since the sensitivity limit is set by
$S_{int}/\theta_{max}\!=\!0.025$ Jy/arcmin (see Section
\ref{sec2:sensitivity-limit}), and a given redshift $S_{int} \propto P$ and 
$\theta_{max} \propto D$, the limit at that redshift follows the
relation $P \propto D$. In Fig. \ref{DS1923:p-d-selected} we have assumed a radio
source spectral index of $-0.8$ to convert flux density into radio power.

At a particular redshift, WENSS can only detect giant sources which are more powerful than the radio power at which the line has been drawn (i.e only in that part of the $P-D$ diagram which is situated above the line). 
Note that lines of higher redshifts also start at a larger
linear size because of the $5\arcmin$ lower angular size limit we have
imposed. Based on the accessible regions in the $P-D$ diagram there is no apparent reason why sources larger than 2 Mpc should be missed.

\begin{figure}
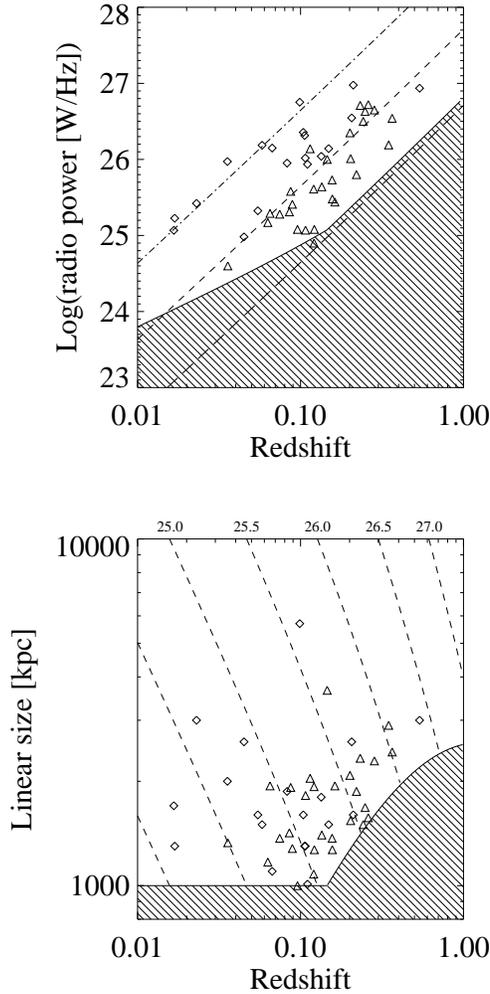

\resizebox{0.8\hsize}{!}{\epsfig{file=DS1923.5a}}\\
\resizebox{0.8\hsize}{!}{\epsfig{file=DS1923.5b}}
\caption{\label{DS1923:p-d-z_diagram} {\bf a}: (Top) The 325-MHz radio power
against the redshift of the formerly known (diamonds) and
newly discovered (triangles) GRGs.  The three diagonal lines represent
the radio power for sources with a 325-MHz flux density of 0.1 Jy
(long-dashed), 1.0 Jy (dashed) and 10.0 Jy (dot-dashed), assuming a
spectral index of $-0.8$. The hatched area indicates the part of the
diagram in which sources fall below our sensitivity limit.  {\bf b}: (Bottom)
The projected linear sizes of the GRGs against their redshift.  The
meaning of the symbols is the same as in the upper plot. The hatched
area indicates the 
region of the plot in which sources have not been selected, either because
they are physically smaller than 1 Mpc, or spatially smaller than
$5\arcmin$. 
The dashed lines indicate the sensitivity limit
for sources of the particular radio power is indicated (in logarithmic
units of W\,Hz$^{-1}$ at 325 MHz) on top of each line. We assume a spectral
index of $-0.8$\,. Sources can only be
detected if they lie to the left of the line belonging to their
particular radio power.}
\end{figure}   

\subsection{Radio power and linear size versus redshift}

In Fig. \ref{DS1923:p-d-z_diagram}a we have plotted the 325-MHz radio
powers of the GRGs against their redshifts. The
higher sensitivity of the WENSS, as compared to earlier surveys, has
enabled the discovery of GRGs with $5-10$ times lower radio power. 
The hatched region in the figure indicates the part of the
parameter space which is not accessible by WENSS, as a result of its limited sensitivity and lower angular size limit. 
Again, in determining this region we have assumed a spectral
index of $-0.8$ for the radio sources, but the
upper edge of the region is not very sensitive to the spectral index.

The broken shape of the upper edge of the hatched region can be understood as
follows. At all redshifts, the WENSS is most sensitive to physically,
and thus spatially, small sources.  Since both a lower
angular and physical size limit have been imposed, there are two regimes in which each of
these two limits is in effect. At very low redshifts, 
a source with a physical size of 1 Mpc extends over
$5\arcmin$ on the sky. This leads to a surface brightness limit
and the integrated flux density of the source thus has to be large for the source to be still detectable.  At higher
redshift, the selection is constrained by the $5\arcmin$ lower
angular size limit. This results in a lower limit to the flux
density of a detectable source, determined by $S_{int} \ge 0.025~[{\rm
Jy/arcmin}] \cdot 5~[{\rm arcmin}] = 0.125$ Jy. Thus, we
are flux density limited. The break occurs 
at that redshift at which a 1-Mpc large radio source will span
an angle of $5\arcmin$~ on the sky ($z\!\sim\!0.146$).

The one source that just lies in the hatched area where
no sources should have been found is B\,1044$+$745, which, as mentioned
before, would indeed not have been selected on basis of the WENSS data. 

The strong effect that the lower angular size limit has on the
selection can best be seen in Fig. \ref{DS1923:p-d-z_diagram}b, where we
have plotted the projected linear size of the identified GRGs against their
redshift. Again, the hatched area indicates the region of the diagram which is
inaccessible. The strong `bump' at redshifts above $\sim 0.15$ 
result from the lower angular size limit.

Furthermore, in Fig. \ref{DS1923:p-d-z_diagram}b we have plotted the sensitivity limit for sources of constant radio power (dashed lines). 
The (logarithm of the) radio power (in W\,Hz$^{-1}$ at 325 MHz) for each particular line has been indicated at the top of the diagram. 
A radio source of radio power $P$ and linear size $D$ can only be selected if it has redshift below that indicated by the dashed line for power $P$ at size $D$. Had that source been at a
higher redshift, or had it been of larger linear size, it would not
have been selected. Likewise, all sources which lie on the left of such
a line of constant $P$, should all have a radio power below this
value. Apart from the before-mentioned case of B\,1044$+$745, this is
indeed the case for the identified GRGs in our sample.

\subsection{The 2-Mpc linear size cut-off}
\label{sec2:cutoff}

A strong drop in the number of sources with projected linear
size above 2 Mpc has been found (see Fig. \ref{DS1923:distributions}c). This may
be a result of a strong negative radio power evolution of radio sources
with increasing size, combined with our sensitivity limit.   
A negative radio power evolution is indeed expected for active radio radio
sources (cf. Kaiser et 
al. \cite{kaiser97}, Blundell et al. \cite{blundell99}). On the other hand, the observed effect
can also be caused if a substantial fraction of the GRGs stop their
radio activity (and `fade') before they reach a size of 3 Mpc.

Above, we have argued that the selection effects alone provide no 
apparent reason why sources of linear size above 2 Mpc
should have been missed if they existed in large numbers. 
On the contrary, Fig. \ref{DS1923:p-d-z_diagram}b)
shows that sources above 2 Mpc in size can potentially be selected out to much
higher redshift than smaller sized sources. 
Indeed, the figure shows that the majority of identified $2-3$~Mpc large
sources have redshifts at which 1-Mpc large sources would not have been
selected. 

Therefore, the observed 2-Mpc cut-off must be caused by a combination of the
internal luminosity evolution of the sources and the sensitivity of the WENSS. 
An extreme case of such a luminosity evolution occurs when a large fraction
of giant sources do not remain active for a long enough amount of time to
reach a linear size of 2 Mpc. To disentangle the effects of the luminosity
evolution of active and so-called `relic' sources on the observed number
of sources as a function of linear size requires much better statistics on the 
death-rate of radio galaxies as a function of radio power and linear
size.


\section*{Acknowledgments}

The INT and WHT are operated on the island of La Palma by the Isaac Newton
Group in the Spanish Observatorio del Roque de los Muchachos of the
Instituto de Astrofisica de Canarias.  The Westerbork Synthesis Radio
Telescope (WSRT) is operated by the Netherlands Foundation for
Research in Astronomy (NFRA) with financial support of the Netherlands
Organization for Scientific Research (NWO).  The National Radio
Astronomy Observatory (NRAO) is operated by Associated Universities,
Inc., and is a facility of the National Science Foundation (NSF).
This research has made use of the NASA/IPAC Extragalactic Database
(NED) which is operated by the Jet Propulsion Laboratory, California
Institute of Technology, under contract with the National Aeronautics
and Space Administration.
The Digitized Sky Surveys were produced at the Space Telescope Science
Institute under U.S. Government grant NAG W-2166. The images of these
surveys are based on photographic data obtained using the Oschin
Schmidt Telescope on Palomar Mountain and the UK Schmidt Telescope. The
plates were processed into the present compressed digital form with the
permission of these institutions.  
M.N. Bremer, H. Sangheera and D. Dallacasa are thanked for their help in
the early stages of this project. P. Best and M. Lehnert are thanked for
many helpful discussions and suggestions. L. Lara is acknowledged fro
providing high resolution radio maps of several sources prior to
publication.  



\begin{appendix}

\section{The sample of newly discovered GRGs and their properties}

Here, we present notes on the radio structure, the host galaxies and
the optical spectra of the sources.  
Tab. \ref{DS1923:radiocores} presents the flux densities and
positions of the radio cores and the host galaxies.
Tab. \ref{DS1923:radioproperties} lists several other radio properties
and the redshifts of the sources. Further, we present a table with the
log of the INT observations (Tab. \ref{DS1923:INT_observations}) and a
table with the wavelengths and redshifts of spectral emission and
absorption features we have measured in the spectra of the sources
(Tab. \ref{DS1923:redshifts}).  Finally, we present the radio and optical
images of the sources, and the optical spectra of their host
galaxies. Radio contour plots mostly are from the WENSS and/or the
NVSS surveys. The FWHM beam size is indicated in a separate panel,
usually situated in the lower left corner of the contour plot. The
flux density level of the first contour can be found in
Tab. \ref{DS1923:figure_captions}.  Unless indicated otherwise in this
table, contour levels have been plotted at
$(-1,1,2,4,8,16,32,64,128,256)$ times the flux density level of the
first contour. For the 1.4-GHz WSRT observations,
Tab. \ref{DS1923:figure_captions} also gives the major and minor axes
(FWHM) of the restoring beam.  Also, we present overlays of our
highest resolution radio map with the optical fields, usually retrieved
from the digitized POSS-I survey, or POSS-II whenever available (see
Tab. \ref{DS1923:figure_captions}).  The range of grey scales is such that
the contents of the optical field is shown best. In case the
identification of the host galaxy of the radio source is not apparent
from the optical image, we have encircled it or used an arrow to
point it out.  The optical spectra of the host galaxies have the
identified emission and absorption features indicated.  In the plots
of the spectra, the `$\oplus$'-symbol indicates
the presence of an atmospheric feature, such as an imperfectly subtracted
bright sky-line or an atmospheric absorption band.  The
`$\downarrow$'-symbol indicates the presence of a feature resulting
from a cosmic ray impact that could not be properly removed from the
data.

\begin{description}
\item{\sl B\,0211+326}: This $\sim\!5\arcmin$-large radio source has been
shortly observed with the WSRT, but no radio core has been detected. 
On the POSS-II image we find a small group of faint galaxies situated
halfway between the two radio lobes. The brightest of these is 
most likely the host galaxy, since its optical spectrum 
shows bright emission lines. We find a redshift of
$0.2605$, so that the projected linear size of the radio
source is 1.6~Mpc.\\

\item{\sl B\,0217+367}: An overlay of a 1.4-GHz WSRT radio map with the optical
field shows that a $\sim 10$ mag. galaxy coincides with a central compact
radio structure. A higher resolution 2.7-GHz map
presented by Faulkner (\cite{faulkner85}) shows a radio core with two jets.
The western jet is bend and has a bright knot at a distance of $20\arcsec$
from the core. Only the western radio lobe is connected to this radio 
core by a faint bridge. The lobe itself shows several twists and a diffuse
outer structure which is only barely detected in the NVSS. The optical 
spectrum of the host galaxy shows weak $[$N{\sc ii}$]$6584 and
$[$S{\sc ii}$]$6717/6731 emission lines.  The redshift of the host
galaxy is 0.0368, yielding a projected linear size of 0.95 Mpc.\\

\item{\sl B\,0648+733}: The compact radio
source located between the two radio lobes coincides with a galaxy on the
POSS-I plates. A more sensitive VLA observation shows that the
western lobe has an edge-brightened radio structure (Lara et al. \cite{lara01});
our WSRT map only shows the bright hotspot. The optical 
spectrum of the identification shows strong emission lines.  The
redshift is $0.1145$, yielding a projected linear size of 2.0
Mpc. The compact source overlapping with the eastern radio lobe on the
NVSS radio map is most likely an unrelated background source.\\

\item{\sl B\,0648+431}: This $\sim\!10\arcmin$-large radio source
consists of four main components. An overlay of the 1.4-GHz
WSRT radio map with an optical image shows that the eastern of the two middle
component coincides with a 12 mag. optical galaxy. An optical spectrum
of the galaxy shows weak $[$N{\sc ii}$]$6548,6583 line-emission on top
of a stellar continuum, at a redshift of $0.0891$. The projected
linear size of the radio source is $1.3$ Mpc.\\

\item{\sl B\,0658+490}: In the WENSS radio map this source is an
$\sim\!19\arcmin$-large complicated radio structure with a fairly
diffuse eastern part, three compact sources in the middle and a diffuse
western extension. The FIRST survey has resolved the central source, and
shows that it has a jet-like extension in the direction of a larger jet-like
feature on the NVSS radio map.  The central source coincides with an 11
mag. Optical galaxy, whose optical spectrum shows weak $[$NII$]$6548/6583 
emission. The redshift is $0.0650$.  The two bright compact sources
west of the radio core are resolved in the FIRST survey. The
westernmost one appears to be an edge-brightened radio lobe, the other
is more compact and has a radio tail pointing away from the radio
core. It is unclear if this component is truly associated with the
large radio source. If it were an unrelated source, the lack of an
optical identification on the POSS-II survey suggests that it is at a
much higher redshift than the host galaxy of the central radio
source. Assuming that all radio structures are part of a single radio
source, it has a linear size of 1.9~Mpc.\\

\item{\sl B\,0747+426}: This is a $6\arcmin$-large radio source with a
bent radio structure. The southern lobe is larger and
more diffuse. Neither of the two lobes contains a bright hotspot.
Near the central bent a 17 mag. galaxy is 
detected which coincides with a weak, unresolved source in the FIRST
survey, presumably the radio core. Its optical spectrum reveals $[$O{\sc ii}$]$3727 emission and a stellar continuum. The redshift is $0.2030$, yielding a projected linear size of 1.5 Mpc.\\

\item{\sl B\,0750+434}: Although only a weak radio source in the
WENSS, this object is one of the largest radio sources we have
discovered in this project. The central source, unresolved by FIRST,
coincides with a 16.7 mag. star-like object. An optical spectrum
reveals strong emission lines, among which broad Hydrogen Balmer
lines, and a blue-colored continuum. The redshift is 0.3474, which is
the highest of all sources presented here. The southern radio lobe is
resolved by FIRST into three separate components, of which probably
only the most southern one is related to the large scale radio
structure. The projected linear size of this source is 2.9 Mpc.\\

\item{\sl B\,0757+477}: The NVSS reveals a strong radio core in this
source. The FIRST survey radio map shows that the northern lobe has a
compact hotspot and that the southern lobe is diffuse. The core is
unresolved by FIRST and coincides with a 15.5 mag. star-like
object. The optical spectrum is dominated by strong and broad Hydrogen
Balmer lines and a blue non-thermal continuum. The redshift is 0.1567,
resulting in a projected linear size of the radio source of 1.3
Mpc. \\

\item{\sl B\,0801+741}: The 1.4-GHz WSRT observations of this source
reveal a compact core, coinciding with a 15.8 mag. optical galaxy.
The radio lobes are largely resolved out, however. The NVSS map
suggests that these lobes are narrow and straight. The redshift of the
host galaxy is 0.1204, and its sepctrum reveals H$\alpha$, $[$N{\sc
ii}$]$6548/6583 and $[$S{\sc ii}$]$6717/6738 emission lines. The
H$\beta$ emission line is weak, relative to H$\alpha$, suggesting a
large amount of extinction towards the line-emitting gas. The
projected linear size of the radio source is 1.1 Mpc.\\

\item{\sl B\,0809+454}: In the WENSS and NVSS this source is a
rather weak $\sim\!7\arcmin$-large radio source. The FIRST survey
radio map shows the radio core and the two outer extremities of the
radio lobes which contain hotspots. The radio core coincides with a
18.5 mag. optical source, whose spectrum shows strong $[$O{\sc
iii}$]$4959,5007 and $[$O{\sc ii}$]$3727 emission lines. The redshift
is 0.2204, yielding a projected linear size of 1.9 Mpc.\\ 

\item{\sl B\,0813+758}: This source has originally been selected from the
NVSS, in which it is an $8.5'$-large radio source with a bright
unresolved central source and an asymmetrical radio structure. A short
1.4-GHz WSRT observation shows that the the eastern radio lobe has a
rather strange structure.  Higher quality VLA data confirm that all
structures on the WSRT map are part of a single large radio lobe (Lara
et al. \cite{lara01}). The central radio source coincides with a
17.5 mag. galaxy whose optical spectrum shows strong emission
lines. The H$\alpha$-line has a broad component. We measure a redshift
of $0.2324$, resulting in a linear size of the radio source of
2.3~Mpc.\\

\item{\sl B\,0905+352}: This $6.3\arcmin$-large radio source shows no
significant structure in WENSS or NVSS. FIRST, however, reveals that
the two radio lobes have strong hotspots, and that the western lobe
has an extension towards the north-west. No radio core is detected.
Two bright compact objects halfway the two radio lobes were identified
as stars, and only the galaxy surrounded by the dashed circle in
Fig. \ref{DS1923:0905+352} was found to show a galaxy-like spectrum,
although we have not yet firmly established its redshift. The two
lines indicated as $[$N{\sc ii}$]$ in the spectrum 
coincide with atmospheric OH-bands. If they
are real, the redshift is 0.106, which is rather low considering
that the galaxy has a POSS-E magnitude of 18.0\,. A redshift of
$0.260$ can also be argued, because of a possible 4000\AA-break
observed at a wavelength of 5050\AA. 
This higher redshift would agree more closely with the
optical magnitude of the galaxy.\\

\item{\sl B\,0925+420}: This source is one of the so-called
`Double-double' radio galaxies (see Schoenmakers et al. \cite{schoenmakers00a}), consisting
of two separate double-lobed radio structures. Radio maps, optical
images and further information are presented in Schoenmakers et
al. (\cite{schoenmakers00a}).\\

\item{\sl B\,0935+743}: This source has a radio morphology much
resembling that of the known giant radio galaxy 4CT\,74.17 (e.g. van
Breugel \& Willis \cite{breugel81}). The central source coincides with an optical
galaxy of 14.3 mag.  The optical spectrum shows no signs of emission
lines.  A somewhat higher resolution observation shows that the
central source has two jet-like features pointing towards the outer
two sources (Lara et al. \cite{lara01}), which suggests that all
three are part of a single radio structure. The redshift of the
optical galaxy associated with the radio core is 0.1215, yielding a
projected linear size of 1.3 Mpc.\\

\item{\sl B\,1029+281}: Because of the low declination of this source,
the WENSS radio map does not show much structure. The NVSS map shows a
large double-lobed radio source with a relatively strong radio
core. The central source is resolved by FIRST into a compact source
coinciding with a 14.3 mag. optical galaxy, and a radio lobe-like
feature at a distance of $\sim\!1\arcmin$ to the north. This much
resembles the structure observed in the so-called `double-double'
radio galaxies (Schoenmakers et al. \cite{schoenmakers00a}). The optical
spectrum of the host galaxy shows strong and broad H$\alpha$ emission.
The H$\beta$ emission line also has a broad component. Also, it is
much weaker than H$\alpha$ which may indicate a high amount of
internal extinction. The redshift is 0.0854, yielding a projected
linear size of 1.4 Mpc.\\

\item{\sl B\,1044+745}: Although this radio source is clearly detected in
the NVSS, it is not in the WENSS.  The NVSS radio map shows two
diffuse lobe-like structures with a weak, compact source in
between. The compact source coincides with a 14 mag. galaxy. The
optical spectrum of this galaxy is dominated by stellar continuum,
with only a weak $[$N{\sc ii}$]$6584 emission-line. The redshift is
0.1210, and the projected linear size of the radio source is thus 1.9
Mpc.\\

\item{\sl B\,1110+405}: A FIRST radio map of this $12\arcmin$-large
radio source 
shows a relatively bright component in the western radio lobe which
resembles a small edge-brightened radio lobe. However, there also is
diffuse radio emission at much a larger distance from the radio
core. The radio core is identified with a 11.4 mag. galaxy, which
appears to be double. However, the western optical object is
spectroscopically identified as a star. The optical spectrum of the
galaxy shows stellar continuum and weak $[$N{\sc ii}$]$6584 and
$[$S{\sc ii}$]$6717/6731 emission lines although the latter lie in an
atmospheric absorption band. The projected linear size of the radio
source is 1.4 Mpc.\\

\item{\sl B\,1144+352}: This source is extensively discussed elsewhere
(Schoenmakers et al. \cite{schoenmakers99b}).\\

\item{\sl B\,1213+422}: The radio map from the FIRST survey shows two
extended radio lobes and an unresolved central radio source, which
coincides with a 15.9 mag. galaxy. Its optical spectrum shows strong
$[$O{\sc iii}$]$4959,5007 emission lines and stellar continuum. The
H$\alpha$-line is clearly broadened. The redshift is $0.2426$ and the
projected linear size is 1.5~Mpc.\\

\item{\sl B\,1245+676}: This $\sim\!12\arcmin$-large radio source has an
inverted spectrum radio core, which is identified with a 14
mag. galaxy. An optical spectrum of this source can be found in Marcha
et al. (\cite{marcha96}); it shows no sign of emission lines. The redshift is
0.1073, and the projected linear size of the radio source is therefore
1.8 Mpc. A more detailed analysis of the radio structure of this
source will be presented in de Bruyn et al. (in preparation).\\

\item{\sl B\,1306+621}: This $\sim\!9\arcmin$-large radio source is rather faint.  
The WENSS and
NVSS radio maps show two radio lobes and a radio core. A 1.4-GHz WSRT
observation shows that both lobes are edge-brightened. The
radio core coincides with a 16 mag. galaxy, which does
not show any emission-lines, though. The redshift is 0.1625, yielding
a projected linear size of 1.9 Mpc.\\ 

\item{\sl B\,1310+451}: On an angular scale, this is the largest radio
source we have found. It has a size of $\sim\!23\arcmin$ and
contains a bright central core and two diffuse outer lobes. The bright
compact source just beyond the western lobe is most likely an
unrelated radio source. In the FIRST survey, the radio core is
resolved and two jets are visible. The radio core coincides with a 10
mag. galaxy. The optical spectrum shows that it has weak $[$N{\sc
ii}$]$6584 and $[$S{\sc ii}$]$6717/6731 emission lines. It has a
redshift of 0.0358, yielding a projected linear size of 1.4 Mpc.
The existence of the large scale radio emission and the redshift of this
radio source have been reported earlier by Faulkner (\cite{faulkner85}).\\

\item{\sl B\,1416+380}: This $\sim\!7\arcmin$-large radio source
resembles a fat double radio source. The northern radio lobe is barely
detected in the NVSS. The integrated spectral index of the whole
source between 325 and 1400 MHz is $-1.46$, which makes this radio
source the steepest spectrum radio source in our sample. The central
component is unresolved in the FIRST and it coincides with a 14.5
mag. galaxy. The optical spectrum of this galaxy shows $[$O{\sc
i}$]$6300, H$\alpha$ and $[$N{\sc ii}$]$6548/6584 emission.  Its
redshift is 0.1350, resulting in a projected linear size of 1.4 Mpc.\\

\item{\sl B\,1426+295}: The eastern lobe of this $\sim\!15'$-large source
is situated near a bright, but unrelated radio source.  The FIRST
radio map shows an unresolved central object, which coincides with a
13 mag. optical galaxy. The optical spectrum of this galaxy shows weak
emission lines of $[$N{\sc ii}$]$6548/6583, H$\alpha$, which is partly
in absorption, and $[$S{\sc ii}$]$6717/6731. The redshift is $0.087$,
resulting in a projected linear size of the radio source of
1.9~Mpc. We remark that spectroscopical observations of the galaxy
situated $\sim\!30\arcsec$ to the north-west of the host galaxy show
that it has the same redshift and is therefore most likely part of the
same group of galaxies as the host galaxy.\\

\item{\sl B\,1450+333}: This source is also a 
`Double-double' radio galaxy. Radio maps, optical
images and further information are presented in Schoenmakers et
al. (\cite{schoenmakers00a}).\\ 

\item{\sl B\,1543+845}: This source has been found in the NVSS
maps, before the WENSS maps of this region of the sky became available.
It shows an $8'$-large double-lobed radio source with a
weak central compact source which coincides with an optical galaxy.
The identification of the compact radio source as the radio core is 
confirmed by a higher resolution VLA radio map (Lara et
al. \cite{lara01}).  
In the optical spectrum, which has been obtained by I.M. Gioia with the 2.2m
telescope of the University of Hawaii on Mauna Kea, we have identified
the $[$O{\sc iii}$]$ 4959/5007 emission lines. 
We measure a redshift of $0.201$, which yields a 
projected linear size of 2.1 Mpc for the radio source.\\

\item{\sl B\,1709+464}: This $10\arcmin$-large radio source has a
diffuse western lobe and a more edge-brightened eastern lobe. The radio core is
clearly detected and a map from the FIRST survey shows that this source has a
small two-sided structure somewhat resembling the larger scale radio
structure. The core coincides with a 10 mag. optical galaxy. The spectrum
reveals weak $[$N{\sc ii}$]$6584 and $[$S{\sc ii}$]$6717/6731 emission
lines. The redshift of the host galaxy is 0.0368, yielding a projected
linear size of 0.59 Mpc. This is therefore not a GRG.\\

\item{\sl B\,1736+375}: This is a $7\arcmin$-large elongated radio
structure which in the NVSS map has a dominant central source.  It
coincides with a 15.9 mag. galaxy, whose optical spectrum reveals
H$\alpha$ and $[$N{\sc ii}$]$ emission-lines, although these are
situated near a strong atmospheric absorption band. The redshift is
0.1562.  The northern radio structure appears to be a small
double-lobed source in the NVSS, and at its center a faint optical
galaxy is situated. We have no spectrum of this source.  The northern
structure may be an unrelated radio source.  If the whole structure
were to be a single radio source, the projected linear size would be
1.4 Mpc.\\

\item{\sl B\,1852+507}: The NVSS map of this $7\arcmin$-large radio
source shows two lobes and a central component which coincides with a
12.8 mag. galaxy. We have identified the bright nearby optical source
as a star. The optical spectrum of the galaxy shows no emission
lines. Its redshift is 0.0958, yielding a linear size of the radio
source of 1.0 Mpc.\\

\item{\sl B\,1911+470}: This $\sim\!6\arcmin$-large radio source has a
dominant central component and a somewhat bent southern radio
lobe. The central component coincides with a 12 mag. optical galaxy,
which has a companion at a distance of $\sim\!30\arcsec$ to the
north-west. The optical spectrum shows a weak $[$N{\sc ii}$]$6584
emission-line, but it overlaps with an atmospheric absorption band. Its
redshift is 0.0548. The companion has the same redshift and they are
therefore most likely also physically close to each other. The
projected linear size of the radio source is 0.54 Mpc, and therefore
this is not a GRG.\\

\item{\sl B\,1918+512}: This $7.3\arcmin$-large radio source has a
FRII-type radio morphology. The eastern `extension' is an unrelated
radio source. A deep 21-cm WSRT observation shows a possibly two-sided
jet in this source and a central radio source, probably the radio
core. A CCD-image obtained by P. Best with the LDSS imaging
spectrograph on the 4.2m WHT telescope on La Palma shows a faint
galaxy close to the core; on the DSS this source is merged with a
nearby star. An optical spectrum shows possible emission lines of
$[$O{\sc ii}$]$3727 and $[$O{\sc iii}$]$5007, as well as several
absorption features, all in agreement with a redshift of $0.284$.  We note
that the star-like object situated in the `gap' between the two radio
lobes is spectroscopically identified as a star. 
If the redshift of the host galaxy is
correct, the linear size of this radio galaxy is 2.3 Mpc.\\

\item{\sl B\,2147+816}: This $18\arcmin$-large FRII-type radio galaxy has
been briefly described before by Saunders (\cite{saunders82}), and more recently
(and more extensively) by Palma et al. (\cite{palma00}). Both the NVSS
and WENSS radio map show the two radio lobes and the central core.
The radio core coincides with a 16.5 mag. galaxy, which is a member of
a small group of three galaxies. The optical spectrum of the
identification shows strong emission lines. Its redshift is 0.1457,
confirming the values given by Saunders (\cite{saunders82}) and Palma et
al. (\cite{palma00}). The projected linear size of the radio source is 3.7 Mpc.\\
\end{description} 

\clearpage

\begin{table*}
\setlength{\tabcolsep}{8pt}
\caption{\label{tab:big_list_of_candis} Table with WENSS selected
candidate GRGs after removing sources identified as non-GRGs on basis of
optical data. Column 1 gives IAU-notation (in B1950.0) of the source name. Columns 2 and 3 give approximate source coordinates in Right Ascension and Declination in B1950.0 coordinates. Column 4 gives the integrated WENSS flux of the source. Column 5 gives the size of the source in arcminutes. Column 6 indicates whether the source is considered a GRG-candidate, for reasons given in column 7.}
\begin{tabular}{l l l l l c p{7.5cm}}
\hline \hline 
\multicolumn{1}{c}{(1)} &\multicolumn{1}{c}{(2)} &\multicolumn{1}{c}{(3)} &\multicolumn{1}{c}{(4)} &\multicolumn{1}{c}{(5)} &\multicolumn{1}{c}{(6)} &\multicolumn{1}{c}{(7)} \\ 
\multicolumn{1}{c}{Source} & \multicolumn{1}{c}{R.A.} & \multicolumn{1}{c}{Dec.} & \multicolumn{1}{c}{$S_{325}$} & \multicolumn{1}{c}{Size} &      \multicolumn{1}{c}{Still} & \multicolumn{1}{c}{Reason/comments} \\
& \multicolumn{1}{c}{$[\,h~~m~~s\,]$} & \multicolumn{1}{c}{$[\,\degr ~~ \arcmin ~~ \arcsec\,]$} & \multicolumn{1}{c}{$[$\,Jy\,$]$} & \multicolumn{1}{c}{$[\,
\arcmin\,]$} & \multicolumn{1}{c}{selected?} & \\
\hline\\
\object{B\,0001$+$342} & 00 01 10 & 34 13 00 & 0.69 & 12 & no & NVSS morphology \\
\object{B\,0023$+$750} & 00 23 15 & 75 00 30 & 0.30 & 5  & yes & NVSS selected,
supported by WENSS\\
\object{B\,0036$+$285} & 00 36 05 & 28 29 00 & 1.54 & 7  & no  & NVSS and 1.4-GHz WSRT morphology \\
\object{B\,0050$+$433} & 00 50 10 & 43 25 00 & 0.59 & 12 & no  & NVSS morphology \\
\object{B\,0058$+$403} & 00 58 05 & 40 19 00 & 1.04 & 5  & no  & NVSS morphology \\       
\object{B\,0119$+$377} & 01 19 10 & 37 44 00 & 2.29 & 6  & no  & NVSS morphology \\ 
\object{B\,0126$+$426} & 01 26 45 & 42 36 00 & 0.79 & 13 & no  & NVSS morphology \\
\object{B\,0141$+$762} & 01 41 25 & 76 13 00 & 0.18 & 6  & yes & NVSS selected,
supported by WENSS\\  
\object{B\,0200$+$457} & 02 00 10 & 45 45 30 & 0.19 & 6  & yes & NVSS morphology\\
\object{B\,0200$+$678} & 02 00 30 & 67 53 00 & 2.15 & 6  & no  & 1.4-GHz WSRT morphology\\
\object{B\,0201$+$317} & 02 01 35 & 31 42 00 & 0.15 & 6  & no  & NVSS morphology \\
\object{B\,0211$+$326} & 02 11 20 & 32 37 00 & 1.59 & 5  & yes & 1.4-GHz WSRT morphology\\
\object{B\,0217$+$367} & 02 17 20 & 36 46 00 & 2.44 & 16 & yes & NVSS morphology\\
\object{B\,0330$+$871} & 03 30 35 & 87 07 00 & 0.11 & 6  & yes & NVSS selected,
supported by WENSS\\  
\object{B\,0334$+$329} & 03 34 30 & 32 59 00 & 0.54 & 6  & yes & NVSS morphology\\
\object{B\,0429$+$293} & 04 29 40 & 29 22 00 & 1.31 & 7  & no  & NVSS morphology \\
\object{B\,0503$+$704} & 05 03 30 & 70 27 00 & 1.01 & 6  & no  & NVSS morphology \\
\object{B\,0603$+$612} & 06 03 00 & 61 15 00 & 1.64 & 5  & yes & NVSS morphology\\
\object{B\,0627$+$721} & 06 27 40 & 72 11 30 & 0.77 & 5  & yes & NVSS and 1.4-GHz WSRT morphology\\
\object{B\,0634$+$515} & 06 34 35 & 51 30 00 & 0.36 & 10 & yes & NVSS morphology\\
\object{B\,0646$+$370} & 06 46 55 & 37 03 00 & 0.47 & 6  & no  & NVSS morphology \\
\object{B\,0648$+$734} & 06 48 15 & 73 23 30 & 2.4  & 13 & yes & NVSS morphology\\
\object{B\,0648$+$431} & 06 48 40 & 43 08 30 & 0.71 & 10  & yes & NVSS morphology\\
\object{B\,0658$+$490} & 06 58 20 & 49 03 30 & 1.01 & 19 & yes & NVSS morphology\\
\object{B\,0713$+$432} & 07 13 45 & 43 14 00 & 0.59 & 7  & yes & NVSS and FIRST morphology\\
\object{B\,0730$+$375} & 07 30 00 & 37 30 00 & 0.52 & 10 & no  & NVSS and FIRST morphology\\
\object{B\,0747$+$426} & 07 47 40 & 42 39 00 & 0.64 & 6  & yes & NVSS/FIRST morphology\\
\object{B\,0750$+$434} & 07 50 40 & 43 24 00 & 0.24 & 8  & yes & NVSS/FIRST morphology\\
\object{B\,0757$+$477} & 07 57 55 & 47 44 30 & 0.25 & 6  & yes & NVSS/FIRST morphology\\
\object{B\,0801$+$741} & 08 01 15 & 74 09 00 & 0.55 & 6  & yes & NVSS selected, supported by WENSS\\
\object{B\,0809$+$454} & 08 09 40 & 45 25 00 & 0.26 & 7  & yes & NVSS and FIRST morphology\\
\object{B\,0813$+$758} & 08 13 40 & 75 48 00 & 2.14 & 8  & yes & NVSS selected, supported by WENSS\\
\object{B\,0817$+$427} & 08 17 45 & 42 43 00 & 1.18 & 9  & no  & NVSS and FIRST morphology\\
\object{B\,0817$+$337} & 08 17 50 & 33 43 00 & 0.26 & 6  & no  & NVSS and FIRST morphology\\
\object{B\,0840$+$513} & 08 40 45 & 51 18 00 & 0.98 & 10 & no  & NVSS and FIRST morphology\\
\object{B\,0853$+$292} & 08 53 00 & 29 16 00 & 3.90 & 15 & no  & NVSS and FIRST morphology\\
\object{B\,0854$+$402} & 08 54 00 & 40 12 00 & 0.35 & 6  & no  & NVSS and FIRST morphology\\
\object{B\,0903$+$783} & 09 03 11 & 78 21 30 & 0.20 & 8  & no  & Western `lobe' overlaps bright spiral galaxy\\
\object{B\,0905$+$352} & 09 05 45 & 35 17 30 & 0.57 & 6  & yes & NVSS and FIRST morphology\\                 
\object{B\,0909$+$353} & 09 09 45 & 35 22 00 & 0.50 & 6  & yes & NVSS and FIRST morphology\\                 
\object{B\,0917$+$307} & 09 17 15 & 30 42 30 & 0.46 & 10 & no  & NVSS and FIRST morphology\\
\object{B\,0925$+$420} & 09 25 55 & 42 00 00 & 0.55 & 7  & yes & NVSS and FIRST morphology\\
\object{B\,0935$+$743} & 09 35 00 & 74 19 00 & 0.19 & 7  & yes & NVSS selected, supported by WENSS\\
\object{B\,0936$+$512} & 09 36 40 & 51 17 30 & 0.94 & 6  & no  & WSRT 1.4-GHz and FIRST morphology\\
\object{B\,1001$+$548} & 10 01 30 & 54 48 30 & 0.52 & 13 & yes & NVSS morphology\\
\object{B\,1029$+$281} & 10 29 25 & 28 11 30 & 0.67 & 11 & yes & NVSS and FIRST morphology\\
\object{B\,1029$+$322} & 10 29 40 & 32 12 00 & 0.64 & 11 & no  & WSRT 1.4-GHz and FIRST morphology\\
\object{B\,1030$+$312} & 10 30 25 & 31 12 00 & 0.87 & 6  & no  & NVSS and FIRST morphology\\
\object{B\,1036$+$632} & 10 36 30 & 63 15 30 & 0.14 & 5  & yes & NVSS morphology\\
\object{B\,1044$+$745} & 10 44 15 & 74 35 30 & 0.12 & 11 & yes & NVSS selected, supported by WENSS\\ 
\object{B\,1054$+$488} & 10 54 10 & 48 52 30 & 0.71 & 8  & yes & NVSS and FIRST morphology\\                  
\hline 
\multicolumn{4}{l}{Continued on next page...} \\
\end{tabular}
\end{table*}
\clearpage

\begin{table*}
\setlength{\tabcolsep}{10pt}
\begin{tabular}{l l l l r c p{6.5cm}}
\multicolumn{4}{l}{...Continued from previous page} \\
\hline \hline 
\multicolumn{1}{c}{(1)} &\multicolumn{1}{c}{(2)} &\multicolumn{1}{c}{(3)}
&\multicolumn{1}{c}{(4)} &\multicolumn{1}{c}{(5)} &\multicolumn{1}{c}{(6)}
&\multicolumn{1}{c}{(7)} \\
\multicolumn{1}{c}{Source} & \multicolumn{1}{c}{R.A.} &
\multicolumn{1}{c}{Dec.} & \multicolumn{1}{c}{$S_{325}$} &
\multicolumn{1}{c}{Size} &      \multicolumn{1}{c}{Still} &
\multicolumn{1}{c}{Reason/comments} \\
& \multicolumn{1}{c}{$[\,h~~m~~s\,]$} & \multicolumn{1}{c}{$[\,\degr ~~ \arcmin ~~ \arcsec\,]$} & \multicolumn{1}{c}{$[$\,Jy\,$]$} & \multicolumn{1}{c}{$[\,
\arcmin\,]$} & \multicolumn{1}{c}{selected?} &\\
\hline \\
\object{B\,1110$+$405} & 11 10 20 & 40 34 00 & 0.82 & 12 & yes & NVSS and FIRST morphology\\                  
\object{B\,1112$+$333} & 11 12 15 & 33 19 00 & 3.06 & 6  & no  & WSRT 1.4-GHz and FIRST morphology\\
\object{B\,1144$+$353} & 11 44 45 & 35 18 30 & 0.97 & 12 & yes & $z=0.063$ (NED), so
$D \sim 1.2$ Mpc \\
\object{B\,1150$+$312} & 11 50 50 & 31 12 00 & 0.28 & 9  & no  & NVSS and FIRST morphology\\
\object{B\,1209$+$614} & 12 09 30 & 61 04 00 & 2.54 & 6  & no  & 4C\,61.25, NVSS morphology\\
\object{B\,1213$+$422} & 12 13 40 & 42 16 00 & 1.10 & 5  & yes & NVSS and FIRST morphology\\
\object{B\,1218$+$639} & 12 18 30 & 63 56 30 & 1.19 & 8  & no  & NVSS morphology\\
\object{B\,1232$+$535} & 12 32 40 & 53 35 00 & 0.36 & 10 & yes & NVSS morphology\\
\object{B\,1234$+$836} & 12 34 00 & 83 39 30 & 0.07 & 6  & yes & NVSS selected, supported by WENSS \\
\object{B\,1245$+$676} & 12 45 30 & 67 39 00 & 0.20 & 12 & yes & NVSS selected, $z =
0.1073$ (NED), so $D\sim1.8$ Mpc\\
\object{B\,1250$+$452} & 12 50 45 & 45 17 00 & 1.33 & 10 & no  & NVSS morphology and bright optical ID \\
\object{B\,1306$+$621} & 13 06 50 & 62 10 00 & 0.19 & 9 & yes & NVSS morphology\\
\object{B\,1310$+$451} & 13 10 05 & 45 09 30 & 0.65 & 23 & yes & $z=0.0356$ (NED), so $D
\sim 1.4$ Mpc\\
\object{B\,1330$+$361} & 13 30 20 & 36 09 00 & 0.27 & 7  & no  & NVSS and FIRST morphology\\
\object{B\,1340$+$382} & 13 40 50 & 38 15 00 & 0.34 & 10 & yes & NVSS and FIRST morphology\\                  
\object{B\,1340$+$447} & 13 40 55 & 44 44 00 & 0.30 & 6  & no  & NVSS and FIRST morphology\\
\object{B\,1342$+$407} & 13 42 30 & 40 44 00 & 0.51 & 8  & no  & NVSS and FIRST morphology\\
\object{B\,1343$+$371} & 13 43 45 & 37 07 00 & 1.90 & 7  & no  & WSRT 1.4-GHz and FIRST morphology\\
\object{B\,1404$+$362} & 14 04 55 & 36 15 00 & 0.24 & 6  & no  & NVSS and FIRST morphology\\
\object{B\,1415$+$685} & 14 15 30 & 68 33 00 & 0.26 & 6  & yes & NVSS morphology\\
\object{B\,1416$+$380} & 14 16 35 & 38 00 00 & 0.46 & 7  & yes & NVSS and FIRST morphology\\                  
\object{B\,1426$+$295} & 14 26 10 & 29 32 00 & 1.10 & 15 & yes & NVSS and FIRST morphology\\                  
\object{B\,1443$+$310} & 14 43 20 & 31 04 00 & 0.41 & 7  & yes & NVSS and FIRST morphology\\                  
\object{B\,1450$+$333} & 14 50 55 & 33 21 00 & 1.38 & 6  & yes & NVSS and FIRST morphology\\                  
\object{B\,1532$+$315} & 15 32 40 & 31 35 00 & 0.73 & 13 & no  & NVSS and FIRST morphology\\
\object{B\,1535$+$613} & 15 35 40 & 61 22 00 & 0.09 & 6  & no  & NVSS morphology\\
\object{B\,1543$+$845} & 15 43 55 & 84 33 00 & 1.14 & 8  & yes & NVSS selected, supported by WENSS\\
\object{B\,1614$+$485} & 16 14 25 & 48 33 00 & 0.17 & 10 & yes & NVSS and FIRST morphology\\
\object{B\,1623$+$410} & 16 23 30 & 41 03 00 & 1.25 & 7  & no  & NVSS and FIRST morphology\\
\object{B\,1634$+$503} & 16 34 15 & 50 23 00 & 0.59 & 7  & no  & NVSS and FIRST morphology\\
\object{B\,1637$+$539} & 16 37 50 & 53 55 00 & 1.99 & 9  & no  & NVSS and FIRST morphology\\
\object{B\,1639$+$328} & 16 39 05 & 32 50 00 & 0.90 & 6  & no  & NVSS and FIRST morphology\\
\object{B\,1709$+$464} & 17 09 30 & 46 28 00 & 1.11 & 10 & yes & NVSS and FIRST morphology\\
\object{B\,1736$+$375} & 17 36 40 & 37 35 00 & 0.48 & 7  & yes & NVSS morphology\\
\object{B\,1838$+$658} & 18 38 05 & 65 52 00 & 0.95 & 7  & yes & NVSS; $z=0.23$ (NED), so $D \sim 1.9$ Mpc. \\
\object{B\,1844$+$653} & 18 44 00 & 65 19 10 & 0.26 & 7  & yes & NVSS morphology\\
\object{B\,1852$+$507} & 18 52 20 & 50 42 00 & 0.27 & 7  & yes & NVSS morphology\\
\object{B\,1855$+$310} & 18 55 20 & 31 00 00 & 1.35 & 9  & yes & NVSS morphology\\
\object{B\,1911$+$470} & 19 11 50 & 47 01 00 & 0.80 & 6  & yes & NVSS morphology\\
\object{B\,1911$+$481} & 19 11 50 & 48 09 00 & 0.77 & 16 & yes & NVSS morphology\\ 
\object{B\,1918$+$453} & 19 18 40 & 45 22 00 & 0.78 & 18 & yes & NVSS morphology\\ 
\object{B\,1918$+$516} & 19 18 05 & 51 36 00 & 1.22 & 7  & yes & NVSS morphology\\ 
\object{B\,1919$+$479} & 19 19 55 & 47 59 30 & 3.52 & 10 & yes & 4C47.51, $z=0.102$ (NED), so $D\sim1.5$ Mpc\\
\object{B\,1919$+$741} & 19 19 15 & 74 09 30 & 2.04 & 6  & yes & NVSS morphology\\ 
\object{B\,1924$+$549} & 19 24 30 & 54 59 00 & 0.48 & 7  & no  & NVSS morphology\\
\object{B\,2130$+$341} & 21 30 05 & 34 07 00 & 0.37 & 6  & yes & NVSS morphology\\ 
\object{B\,2147$+$816} & 21 47 20 & 81 41 00 & 1.06 & 18 & yes & NVSS selected, supported by WENSS\\
\object{B\,2205$+$376} & 22 05 25 & 37 36 00 & 0.32 & 7  & yes & NVSS morphology\\  
\object{B\,2231$+$320} & 22 31 10 & 32 00 00 & 0.29 & 11 & no  & NVSS morphology\\
\object{B\,2233$+$373} & 22 33 20 & 37 20 00 & 0.62 & 6  & yes & NVSS morphology\\
\object{B\,2312$+$419} & 23 12 40 & 41 57 10 & 0.27 & 6  & no  & NVSS morphology\\
\object{B\,2315$+$401} & 23 15 55 & 40 10 17 & 0.11 & 6  & yes & NVSS morphology\\
\object{B\,2326$+$315} & 23 26 00 & 31 35 00 & 0.47 & 10 & no  & NVSS morphology\\
\object{B\,2357$+$401} & 23 57 10 & 40 08 00 & 0.22 & 5  & yes & NVSS morphology\\
\hline \\
\end{tabular}
\end{table*}
\clearpage

\begin{table*}[t]
\centering
\setlength{\tabcolsep}{12pt}
\caption{\label{DS1923:INT_observations}
Log of the spectroscopic observations of GRG candidates in our sample.
Column 1 gives the name of the source in IAU format. Column 2 gives
the telescope used for the observations. Column 3 gives the observing
date. Column 4 gives the central wavelength of the observation in
{\AA}ngstrom (for the
INT only). Column 5 gives the used width of the slit in arcsec. Column 6
gives the integration time. Column 6 gives an indication of the observing conditions; `P' stands for photometric conditions, `NP' for non-photometric conditions, or cirrus clouds, and `C' stands for cloudy conditions.}
\begin{tabular}{l c l c c r l}
\hline \hline \\ 
\multicolumn{1}{c}{(1)} &\multicolumn{1}{c}{(2)} &\multicolumn{1}{c}{(3)} &\multicolumn{1}{c}{(4)} &\multicolumn{1}{c}{(5)} &\multicolumn{1}{c}{(6)} & 
\multicolumn{1}{c}{(7)}\\ 
\multicolumn{1}{c}{Source} & Tel. & \multicolumn{1}{c}{Date} & $\lambda_{\rm central}$ & Slit & Int. & \multicolumn{1}{c}{Cond.}\\
& & & $[\,$\AA$\,]$ & $[\,\arcsec\,]$ & $[\,$s$\,]$ & \\
\hline \\
B\,0211$+$326 & INT & 4 Aug. 1995 & 6000 & 2 & 2400 & NP \\
         &     & 8 Oct. 1996 & 6500 & 2 & 1200 & P \\
B\,0217$+$367 & INT & 4 Aug. 1995 & 5500 & 2 & 900  & C \\
B\,0648$+$733 & INT & 6 Apr. 1996 & 6000 & 2 & 1800 & P \\
B\,0648$+$431 & INT & 7 Apr. 1996 & 6000 & 2 & 1200 & NP \\
B\,0658$+$490 & INT & 6 Apr. 1996 & 6000 & 2 & 1800 & P\\ 
B\,0747$+$426 & INT & 8 Oct. 1996 & 6000 & 2 & 1200 & P\\
B\,0750$+$434 & INT & 9 Oct. 1996 & 6000 & 2 & 600  & P\\
         &     &             & 6500 & 2 & 600  & P\\
B\,0757$+$477 & INT & 7 Apr. 1996 & 5500 & 2 & 600  & NP \\
         &     &             & 6500 & 2 & 600  & NP\\
B\,0801$+$741 & INT & 7 Apr. 1996 & 6000 & 2 & 600  & NP\\
B\,0809$+$454 & INT & 8 Apr. 1996 & 6000 & 3 & 600  & NP\\
         &     &             & 6500 & 3 & 600  & NP\\
B\,0813$+$758 & INT & 6 Apr. 1996 & 6000 & 2 & 600  & P\\ 
         &     &             & 6500 & 2 & 1200 & P\\
B\,0905$+$352 & INT & 7 Apr. 1996 & 6000 & 2 & 600  & NP\\
B\,0925$+$420 & INT & 8 Apr. 1996 & 6000 & 2 & 600  & NP\\
B\,0935$+$743 & INT & 7 Apr. 1996 & 6000 & 2 & 1200 & NP\\
B\,1029$+$281 & INT & 7 Apr. 1996 & 6000 & 2 & 1200 & NP\\
B\,1044$+$745 & INT & 7 Apr. 1996 & 6000 & 2 & 600  & NP\\
B\,1110$+$405 & INT & 6 Apr. 1996 & 6000 & 2 & 1200 & P\\
B\,1213$+$422 & INT & 7 Apr. 1996 & 6000 & 2 & 600  & NP\\
         &     &             & 6500 & 2 & 600  & NP \\
B\,1306$+$621 & INT & 8 Apr. 1996 & 6000 & 2 & 1200 & NP \\
B\,1310$+$451 & INT & 5 Aug. 1995 & 5500 & 2 & 900  & P \\
B\,1416$+$380 & INT & 5 Aug. 1995 & 6000 & 2 & 900  & P \\
         &     & 9 Apr. 1996 & 6500 & 2 & 600  & P \\
B\,1426$+$295 & INT & 4 Aug. 1995 & 5500 & 2 & 1800 & P \\
         &     & 9 Apr. 1996 & 6000 & 2 & 1200 & P \\
B\,1450$+$333 & WHT & 8 July 1997 &      & 2 & 600  & P \\
B\,1543$+$845 & 2.2m UH  & 4 Mar. 1998 & & 2 & 2100 & P \\
         &          & 5 Mar. 1998 & & 2 & 2400 & P \\
B\,1709$+$464 & INT & 4 Aug. 1995 & 5500 & 2 & 900  & P \\
B\,1736$+$375 & INT & 9 Oct. 1996 & 6000 & 2 & 1200 & P \\
B\,1852$+$507 & INT & 9 Oct. 1996 & 6000 & 2 & 600  & P \\
B\,1911$+$470 & INT & 8 Oct. 1996 & 6000 & 2 & 1200 & P \\
B\,1918$+$516 & INT & 8 Oct. 1996 & 6000 & 2 & 600  & P \\
B\,2147$+$816 & INT & 9 Oct. 1996 & 5500 & 2 & 600  & P \\
         &     &             & 6000 & 2 & 1200 & P \\
\hline \hline \\
\end{tabular}
\vspace*{\textheight}
\end{table*}
\clearpage

\begin{table*}[t!]
\centering
\setlength{\tabcolsep}{5.8pt}
\caption{\label{DS1923:radiocores} Properties of the radio
cores and the optical identifications of the spectroscopically observed
sources, and of the confirmed giant sources B\,1144+352, B\,1245+676 and
B\,1310+451. Column 1 gives the name of the radio source in IAU notation;
column 2 gives the observation used to determine the radio core
position and its flux density; columns 3 and 4 give the radio core
position in right ascension and declination, respectively, in B1950.0
coordinates. These have been obtained by fitting a Gaussian in the
radio map.  Column 5 gives the integrated flux density at 1.4 GHz of
the radio core. Columns 6 and 7 give the position of the optical
identification in right ascension and declination, respectively, in
B1950.0 coordinates, obtained from fitting a Gaussian in the available
optical image. Column 8 gives the magnitude of the
identification in the red (POSS-E) band of the Palomar survey. 
The magnitudes for sources weaker than 15.0 have been
obtained from the APM catalogue and are estimated to
be accurate to 0.5 mag. For brighter sources, we have measured the
magnitudes directly from the digitized POSS-I frames using the
photometric calibration for stars available from the STScI WWW-pages
and through the {\sc getimage-2.0} plate retrieval software. Typical
uncertainties in these values are estimated to be large, at least 1 mag.}
\begin{tabular}{l l l@{$\,\pm\,$}r l@{$\,\pm\,$}r r@{$\,\pm\,$}l
l@{$\,\pm\,$}r l@{$\,\pm\,$}r r}
\hline \hline \\
 \multicolumn{1}{c}{(1)} & \multicolumn{1}{c}{(2)} & \multicolumn{2}{c}{(3)} & \multicolumn{2}{c}{(4)} & \multicolumn{2}{c}{(5)} & \multicolumn{2}{c}{(6)} & \multicolumn{2}{c}{(7)} & \multicolumn{1}{c}{(8)}\\ 
 & & \multicolumn{4}{c}{\hrulefill~Radio~\hrulefill} & \multicolumn{2}{c}{} &
 \multicolumn{4}{c}{\hrulefill~Optical~\hrulefill} & \\ 
Source & Obs. & \multicolumn{2}{c}{R.A.} & \multicolumn{2}{c}{Dec.} &
\multicolumn{2}{c}{Flux} & \multicolumn{2}{c}{R.A.} &
\multicolumn{2}{c}{Dec.} & Mag.\\
& & \multicolumn{4}{c}{(B1950.0)} & \multicolumn{2}{c}{$[$mJy$]$} &
\multicolumn{4}{c}{(B1950.0)} & \\
\hline \\
B\,0211$+$326      & WSRT &\multicolumn{4}{c}{} & \multicolumn{2}{c}{$<2.7$}
& 02 11 18.11 & 0.02 & 32 37 06.8 & 0.2 & 17.6 \\
B\,0217$+$367        & WSRT & 02 17 21.78 & 0.01 & 36 45 58.1 & 0.1 & 64.1 &
3.3 & 02 17 22.02 & 0.02 & 36 45 56.7 & 0.3 & 10.0\\
B\,0648$+$733        & WSRT &  06 48 14.35  & 0.15 & 73 23 33.1 & 0.8 & 2.3 &
0.2 & 06 48 15.01 & 0.01 & 73 23 32.9 & 0.2 & 15.7 \\
B\,0648$+$431        & WSRT & 06 48 43.34 & 0.02 & 43 08 38.8 & 0.3 & 36.2 &
2.1 & 06 48 43.44 & 0.02 & 43 08 36.4 & 0.2 & 11.7\\
B\,0658$+$490        & FIRST & 06 58 18.36  & 0.01 & 49 03 40.9 & 0.1 & 14.2 &
0.5 & 06 58 18.11 & 0.01 & 49 03 37.0 & 0.1 & 10.9 \\
B\,0747$+$426        & FIRST & 07 47 40.54 & 0.03 & 42 39 04.9 & 0.4 & 1.1 &
0.2 & 07 47 40.65 & 0.01 & 42 39 04.0 & 0.1 & 17.0\\
B\,0750$+$434        & FIRST & 07 50 38.63 & 0.01 & 43 24 02.3 & 0.1 & 14.6 &
0.4 & 07 50 38.75 & 0.01 & 43 24 01.1 & 0.1 & 16.7\\
B\,0757$+$477        & FIRST & 07 57 54.64 & 0.01 & 47 44 36.0 & 0.1 & 64.1 &
1.3 & 07 57 54.78 & 0.01 & 47 44 34.5 & 0.1 & 15.5\\
B\,0801$+$741        & WSRT & 08 01 15.88 & 0.03 & 74 09 19.1 & 0.2 & 14.1 &
0.9 & 08 01 16.04 & 0.01 & 74 09 17.7 & 0.1 & 15.8 \\
B\,0809$+$454        & FIRST & 08 09 40.30 & 0.02 & 45 25 05.1 & 0.2 & 1.9 &
0.2 & 08 09 40.41 & 0.01 & 45 25 04.1 & 0.1 & 18.5\\
B\,0813$+$758        & WSRT  & 08 13 39.58  & 0.03 & 75 48 02.7 & 0.1 & 32.9  &
1.9 & 08 13 39.87 & 0.04 & 75 48 00.8 & 0.2 & 17.5\\
B\,0905$+$352        & FIRST & \multicolumn{4}{c}{} &
\multicolumn{2}{c}{$<0.45$} & 09 05 41.90 & 0.03 & 35 18 31.9 & 0.4 & 18.0\\
B\,0925$+$420        & FIRST & 09 25 59.72 & 0.01 & 41 59 54.7 & 0.1 & 6.7 &
0.4 & 09 25 59.95 & 0.01 & 41 59 54.2 & 0.1 & 18.3 \\
B\,0935$+$743        & NVSS & 09 34 56.30 & 0.28 & 74 19 02.7 & 1.4 & 15.6 &
1.2 & 09 34 55.66 & 0.01 & 74 19 03.9 & 0.1 & 14.2 \\
B\,1029$+$281        & FIRST & 10 29 26.77 & 0.01 & 28 11 28.7 & 0.1 & 35.4 &
0.8 & 10 29 26.80 & 0.01 & 28 11 27.2 & 0.1 & 14.3 \\
B\,1044$+$745        & NVSS & 10 44 08.08 & 1.24 & 74 35 19.8 & 6.4 & 4.1 & 1.2
& 10 44 04.53 & 0.02 & 74 35 25.8 & 0.1& 14.0  \\
B\,1110$+$405        & FIRST &  11 10 19.90 & 0.01 & 40 33 49.3 & 0.1 & 10.9 &
0.5 & 11 10 19.92 & 0.01 & 40 33 48.1 & 0.1 & 11.4 \\
B\,1144$+$352        & FIRST &  11 44 45.52 & 0.01 & 35 17 47.4 & 0.1 & 609.3 &
 12.2$^a$ & 11 44 45.55 & 0.01 & 35 17 46.7 & 0.1 & 14.2 \\
B\,1213$+$422        & FIRST & 12 13 39.78 & 0.01 & 42 16 08.2 & 0.1 & 18.2 &
0.5 & 12 13 39.76 & 0.01 & 42 16 06.6 & 0.1 & 15.9 \\
B\,1245$+$676        & NVSS  & 12 45 32.12 & 0.01 & 67 39 37.9 & 0.1 & 269.8 &
 5.4 & 12 45 32.19 & 0.01 & 67 39 37.1 & 0.1 & 14.3 \\  
B\,1306$+$621        & WSRT & 13 06 47.49 & 0.02 & 62 10 14.9 & 0.1 & 9.2 & 0.5
& 13 06 47.66 & 0.01 & 62 10 13.7 & 0.1 & 16.0 \\
B\,1310$+$451        & FIRST & 13 10 03.59 & 0.01 & 45 06 15.2 & 0.1 & 52.8 &
1.1 & 13 10 03.59 & 0.01 & 45 06 15.4 & 0.1 & 9.8\\
B\,1416$+$380        & FIRST & 14 16 33.21 & 0.01 & 38 00 10.3 & 0.1 & 5.2 &
0.4 & 14 16 33.35 & 0.01 & 38 00 09.8 & 0.1 & 14.5\\ 
B\,1426$+$295        & FIRST & 14 26 08.03 & 0.01 & 29 32 06.3 & 0.1 & 12.8 &
0.4 & 14 26 08.09 & 0.01 & 29 32 05.8 & 0.1 & 13.0  \\
B\,1450$+$333        & FIRST & 14 50 58.95 & 0.02 & 33 20 55.0 & 0.2 & 1.9  &
0.2 & 14 50 59.03 & 0.01 & 33 20 53.6 & 0.1 & 18.3 \\
B\,1543$+$845        & NVSS  & 15 43 55.6  & 1.8  & 84 32 52.6 & 2.5 & 5.8  &
0.5 & 15 43 48.93 & 0.02 & 84 32 44.1 & 0.1 & 18.7 \\
B\,1709$+$464        & FIRST & 17 09 33.20 & 0.01 & 46 27 56.5 & 0.1 & 13.2 &
0.4 & 17 09 33.31 & 0.01 & 46 27 56.0 & 0.1 & 10.4\\
B\,1736$+$375        & NVSS & 17 36 38.03 & 0.01 & 37 35 12.3 & 0.1 & 119.0 &
2.5 & 17 36 38.35 & 0.01 & 37 35 11.7 & 0.1 & 15.9 \\
B\,1852$+$507        & NVSS & 18 52 18.02 & 0.05 & 50 42 22.8 & 0.7 & 34.4 &
1.4 & 18 52 18.36 & 0.01 & 50 42 17.0 & 0.1 & 12.8 \\
B\,1911$+$470        & NVSS & 19 11 50.21 & 0.01 & 47 01 43.6 & 0.1 & 164.9 &
3.5 & 19 11 50.52 & 0.01 & 47 01 36.9 & 0.1 & 11.6\\
B\,1918$+$516        & WSRT & 19 18 08.42 & 0.01 & 51 37 51.4 & 0.1 & 19.2 &
0.4 & 19 18 08.55 & 0.01 & 51 37 56.3 & 0.1 & 19.2$^b$\\
B\,2147$+$816        & NVSS & 21 47 12.01 & 0.4 & 81 40 58.3 & 0.9 & 12.9 & 0.9
& 21 47 12.30 & 0.01 & 81 40 57.3 & 0.1 & 16.5 \\ 
\hline \hline \\
\end{tabular}
\begin{minipage}{\linewidth}
Notes:\\
$a-$Variable (see Schoenmakers et al. \cite{schoenmakers99b})\\
$b-$Merged with nearby star on DSS; the magnitude has been determined by
subtracting the flux from the star, obtained by fitting a Gaussian, from
the integrated flux of the star and galaxy combined. The error is therefore large (estimated at 1 mag.).\\
\end{minipage}

\vspace{\textheight}
\end{table*}

\begin{table*}[t!]
\centering
\setlength{\tabcolsep}{8.0pt}
\caption{\label{DS1923:radioproperties} Radio properties of the sources from
Tab. \ref{DS1923:radiocores}. Column 1 gives the source name in IAU
format. Column 2 gives the integrated flux density of the source at
325~MHz from the WENSS (unless states otherwise). Column 3 gives the
integrated flux density at 1400~MHz from the NVSS. Column 5 gives the
spectral index between 325 and 1400 MHz. Column 6 gives the redshift of the
host galaxy. Column 7 gives the angular size of the radio source in
arcminute. Column 8 gives the projected linear size in Mpc. Column 9 gives
the radio luminosity at an emitted frequency of 325 MHz.}
\begin{tabular}{l r@{$\,\pm\,$}l r@{$\,\pm\,$}l l@{$\,\pm\,$}r l@{$\,\pm\,$}r r@{$\,\pm\,$}l
l@{$\,\pm\,$}r l@{$\,\pm\,$}r }
\hline \hline \\
\multicolumn{1}{c}{(1)} & \multicolumn{2}{c}{(2)} &
\multicolumn{2}{c}{(3)} & \multicolumn{2}{c}{(4)} &
\multicolumn{2}{c}{(5)} & \multicolumn{2}{c}{(6)} &\multicolumn{2}{c}{(7)}
& \multicolumn{2}{c}{(8)}\\  
\multicolumn{1}{c}{Source\phantom{11}}  & \multicolumn{2}{c}{$S_{325}$} & \multicolumn{2}{c}{$S_{1400}$}
& \multicolumn{2}{c}{$\alpha_{325}^{1400}$}  & \multicolumn{2}{c}{$z$}  &
\multicolumn{2}{c}{$\theta$} & \multicolumn{2}{c}{$D$}  &
\multicolumn{2}{c}{$\log(P_{325})$} \\
         & \multicolumn{2}{c}{$[$Jy$]$} & \multicolumn{2}{c}{$[$mJy$]$}
& \multicolumn{2}{c}{}  & \multicolumn{2}{c}{}  &
\multicolumn{2}{c}{$[~\arcmin~]$} & \multicolumn{2}{c}{$[$Mpc$]$}  &
\multicolumn{2}{c}{$[$W Hz$^{-1}]$} \\
\hline \\
B\,0211$+$326 & 1.67 & 0.04 & 470 & 11 & $-0.87$ & 0.03 & 0.2605 & 0.0002 & 5.2 & 0.1 & 1.57 & 0.03 & 26.72 & 0.02 \\
B\,0217$+$367 & 2.70 & 0.06 & 828 & 18 & $-0.81$ & 0.03 & 0.0368  & 0.0003  & 15.8 & 0.5 & 0.95 & 0.03 & 25.20 & 0.02 \\
B\,0648$+$733 & 2.41 & 0.06$^a$ & 861 & 18$^b$ & $-0.70$ & 0.04 & 0.1145 & 0.0002 & 12.6 & 0.2 & 1.95 & 0.04 & 26.14 & 0.01 \\  
B\,0648$+$431 & 0.75 & 0.03 & 328 &  7 & $-0.57$ & 0.04 & 0.0891 & 0.0002 & 9.6 &
0.5 & 1.28 & 0.07 & 25.41 & 0.02 \\
B\,0658$+$490 & 1.07 & 0.04 & 399 &  9 & $-0.68$ & 0.04 & 0.0650 & 0.0002 & 19.1 &
0.2 & 1.94 & 0.02 & 25.29 & 0.02\\
B\,0747$+$426 & 0.62 & 0.03$^c$ & 164 & 6$^d$ & $-0.91$ & 0.06 & 0.2030 & 0.0004 &
6.0 & 0.5 & 1.54 & 0.13 & 26.06 & 0.03\\
B\,0750$+$434 & 0.29 & 0.02 & 124 & 4  & $-0.58$ & 0.07 & 0.3474 & 0.0003 & 8.1 &
0.1 & 2.90 & 0.04 &26.19 & 0.03 \\
B\,0757$+$477 & 0.29 & 0.02 & 160 & 3  & $-0.41$ & 0.06 & 0.1567 & 0.0002 & 6.0 &
0.3 & 1.27 & 0.06 & 25.48 & 0.03  \\ 
B\,0801$+$741 & 0.62 & 0.04 & 150 & 4  & $-0.97$ & 0.06 & 0.1204 & 0.0002 & 6.3 &
0.2 & 1.08 & 0.05 & 25.61 & 0.03  \\
B\,0809$+$454 & 0.30 & 0.02 & 125 & 4  & $-0.60$ & 0.06 & 0.2204 & 0.0003 & 6.9 &
0.1 & 1.87 & 0.03 & 25.80 & 0.03  \\
B\,0813$+$758 & 2.14 & 0.05$^c$ & 621 & 13 & $-0.89$ & 0.03 & 0.2324 & 0.0003 &
8.3 & 0.2 & 2.33 & 0.06 & 26.71 & 0.01\\
B\,0905$+$352 & 0.59 & 0.02 & 200 & 5  & $-0.74$ & 0.04 &0.106 & 0.001 & 6.3 &
0.1 & 0.98 & 0.02 & 25.47 & 0.02 \\
\multicolumn{7}{c}{ }  & 0.260 & 0.002 & \multicolumn{2}{c}{ } & 1.90 &
0.03 & 26.26 & 0.02  \\
B\,0925$+$420$^{e}$ & 0.55 & 0.02 & 169 & 4 & $-0.81$ & 0.04 &0.365 & 0.005
& 6.6 & 0.2 & 2.43 & 0.07 & 26.54 & 0.02  \\
B\,0935$+$743 & 0.19 & 0.03 & 94  & 2  & $-0.48$ & 0.11 & 0.1215 & 0.0003 & 7.3  &
0.3 & 1.27 & 0.05 & 25.08 & 0.04 \\
B\,1029$+$281 & 0.65 & 0.05$^f$ & 295 & 7$^f$  & $-0.54$ & 0.07 & 0.0854 & 0.0002
& 11.0 & 0.3 & 1.42 & 0.04 & 25.31 & 0.03  \\
B\,1044$+$745 & 0.12 & 0.06$^g$ & 85  & 5  & $-0.3$  & 0.3  & 0.1210 & 0.0003 &
11.2 & 0.3 & 1.93 & 0.05 & 24.9 & 0.2 \\
B\,1110$+$405 & 0.79 & 0.04$^h$ & 246 & 6  & $-0.80$ & 0.05 & 0.0745 & 0.0003 &
12.0 & 1.0 & 1.37 & 0.11 & 25.28 & 0.03\\
B\,1144$+$352$^i$ & 0.89 & 0.06 & 805 & 18 & $-0.07$ & 0.06 & 0.063$^j$ & 0.001 & 
11.8 & 0.1 & 1.17 & 0.01 & 25.17 & 0.03\\
B\,1213$+$422 & 1.19 & 0.03 & 419 & 9 & $-0.71$ & 0.03 & 0.2426 & 0.0002 & 5.2
& 0.1 & 1.50 & 0.03 & 26.50 & 0.01\\
B\,1245$+$676 & 0.26 & 0.03 & 386 & 9 & $+0.27$ & 0.10 & 0.1073$^k$ & 0.0001 & 
11.6 & 0.2 & 1.82 & 0.03 & 25.07 & 0.05 \\
B\,1306$+$621 & 0.24 & 0.02 & 101 & 4  & $-0.59$ & 0.09 & 0.1625 & 0.0004 & 8.9 &
0.1 & 1.94 & 0.02 & 25.44 & 0.04 \\
B\,1310$+$451 & 0.72 & 0.04 & 305 & 8  & $-0.59$ & 0.05 & 0.0358 & 0.0002 & 22.6 &
0.5 & 1.33 & 0.03 & 24.60 & 0.03 \\
B\,1416$+$380 & 0.49 & 0.02 & 58  & 3  & $-1.46$ & 0.03 & 0.1350 & 0.0002 & 7.4 &
0.5 & 1.40 & 0.09 & 25.64 & 0.02 \\
B\,1426$+$295 & 1.15 & 0.04 & 431 & 10 & $-0.67$ & 0.04 & 0.0870 & 0.0003 & 14.7 & 0.4 & 1.92 & 0.05 & 25.58 & 0.02 \\
B\,1450$+$333$^e$ & 1.51& 0.04 & 460 & 10 & $-0.81$ & 0.04 & 0.249 & 0.001 &
5.7 & 0.1 & 1.68 & 0.03 & 26.63 & 0.01\\
B\,1543$+$845  & 1.14 & 0.03$^g$ & 378 & 8 & $-0.80$ & 0.03 & 0.201 & 0.001
& 7.9 & 0.3 & 2.03 & 0.07 & 26.35 & 0.01\\
B\,1709$+$464 & 1.30 & 0.04 & 461 & 10 & $-0.71$ & 0.04 & 0.0368 & 0.0002 & 9.8 &
0.3 & 0.59 & 0.02 & 24.88 & 0.02 \\
B\,1736$+$375 & 0.51 & 0.02 & 237 & 5  & $-0.52$ & 0.04 & 0.1562 & 0.0003 & 6.5 &
0.3 & 1.37 & 0.06 & 25.73 & 0.02 \\
B\,1852$+$507 & 0.30 & 0.02 & 131 & 4  & $-0.57$ & 0.06 & 0.0958 & 0.0003 & 7.0 &
0.3 & 1.00 & 0.04 & 25.08 & 0.03 \\
B\,1911$+$470 & 0.87 & 0.03 & 341 & 4  & $-0.64$ & 0.03 & 0.0548 & 0.0002 & 6.2 &
0.3 & 0.54 & 0.03 & 25.05 & 0.02 \\
B\,1918$+$516 & 1.20 & 0.03$^l$ & 363 & 9$^m$ & $-0.82$ & 0.03 & 0.284 & 0.001$^n$ & 7.3 & 0.1 & 2.32 & 0.03 & 26.65 & 0.01\\
B\,2147$+$816 & 1.06 & 0.05$^g$ & 414 & 10 & $-0.68$ & 0.05 & 0.1457 & 0.0001 &
18.3 & 0.2 & 3.66 & 0.04 & 26.00 & 0.02 \\
\hline \hline\\
\end{tabular}
\begin{minipage}{\linewidth}
Notes:\\  
$a-$Subtracted $41 \pm 7$ mJy background object at R.A. 06 49 02.7, Dec. 73 25 55.\\
$b-$Subtracted $19.8 \pm 1.1$ mJy background object at R.A. 06 49 02.7, Dec. 73 25 55.\\
$c-$Subtracted $149 \pm 11$ mJy background object at R.A. 07 47 48.2, Dec. 42
39 56.\\
$d-$Subtracted $26 \pm 1$ mJy background object at R.A. 07 47 48.2, Dec. 42
39 56.\\
$e-$See Schoenmakers et al. (\cite{schoenmakers00a}) for radio maps and optical spectrum of this source.\\
$f-$Includes background source at R.A. 10 29 36.3  Dec. 28 13 15.9 (flux
density 10 $\pm$ 1 mJy at 1400 MHz).\\
$g-$WENSS polar cap region (observed frequency 351 MHz).\\
$h-$Subtracted $72 \pm 7$ mJy background object at R.A. 11 10 24.8, Dec. 40
38 03.\\
$i-$See Schoenmakers et al. (\cite{schoenmakers99b}) for radio maps and optical spectrum of this source.\\ 
$j-$Colla et al. (\cite{colla75}).\\
$k-$Marcha et al. (\cite{marcha96}).\\ 
$l-$Subtracted $90\pm7$ mJy background source at R.A. 19 18 17.1, Dec. 51 37
53.\\
$m-$Subtracted $26\pm1$ mJy background source at R.A. 19 18 17.1, Dec. 51 37
53.\\
$n-$Uncertain redshift.\\
\end{minipage}
\vspace*{\textheight}
\end{table*}

\clearpage

\begin{table}[t!]
\setlength{\tabcolsep}{6.5pt}
\caption{\label{DS1923:redshifts} 
The measured wavelengths and resulting redshifts of the most prominent
emission and absorption lines. Column 1 gives the name of the source,
column 2 the used line, column 3 the measured wavelength, i.e. the
position of the peak of the Gaussian used in fitting the line, and column
4 the therefrom derived redshift of that line. The last line for each
source gives the average redshift.} 
\begin{tabular}{l l l l}
\hline \hline \\
\multicolumn{1}{c}{(1)} & \multicolumn{1}{c}{(2)} &\multicolumn{1}{c}{(3)} & \multicolumn{1}{c}{(4)}\\ 
\multicolumn{1}{c}{Source} & \multicolumn{1}{c}{Line} & \multicolumn{1}{c}{$\lambda_{\rm peak}/[$\AA$]$} & \multicolumn{1}{c}{Redshift} \\ 
\hline\\ 
B\,0211$+$326 &   $[$O{\sc ii}$]$3727 & 4697.04 &   0.2603 \\
         &  $[$Ne{\sc iii}$]$    & 4877.16 &   0.2606 \\
         &      H$\beta$         & 6127.18 &   0.2605 \\
         &  $[$O{\sc iii}$]$4959 & 6250.79 &   0.2605 \\
         &  $[$O{\sc iii}$]$5007 & 6311.32 &   0.2605 \\
         &                       &         &   0.2605$\,\pm\,$0.0002 \\
\ \\
B\,0217$+$367 &    Ca{\sc ii}3934 & 4078.60 &   0.0368 \\
         &    Ca{\sc ii}3968 & 4115.06 &   0.0371 \\
         &  G-band           & 4462.41 &   0.0366 \\
         &    Mg-b           & 5364.99 &   0.0367 \\
         &    Na D           & 6110.44 &   0.0369 \\
         &                   &         &   0.0368$\,\pm\,$0.0003 \\
\ \\
B\,0648$+$733 &  G-band               & 4798.18 &   0.1146 \\
         &      H$\beta$         & 5417.93 &   0.1146 \\
         &  $[$O{\sc iii}$]$4959 & 5527.37 &   0.1146 \\
         &  $[$O{\sc iii}$]$5007 & 5580.97 &   0.1146 \\
         &    Mg-b               & 5766.97 &   0.1144 \\
         &    $[$O{\sc i}$]$6300 & 7020.84 &   0.1144 \\
         &   $[$N{\sc ii}$]$6583 & 7336.42 &   0.1144 \\
         &                       &         &   0.1145$\,\pm\,$0.0002 \\
\ \\
B\,0648$+$431 &  G-band & 4688.06 &   0.0890 \\
         &    Mg-b & 5636.05 &   0.0891 \\
         &    Na D & 6418.92 &   0.0892 \\
         &         &         &   0.0891$\,\pm\,$0.0002 \\
\ \\
B\,0658$+$490 &    Mg-b               & 5511.23 &   0.0650 \\
         &    Na D               & 6275.81 &   0.0650 \\
         &   $[$N{\sc ii}$]$6583 & 7010.76 &   0.0650 \\
         &                       &         &   0.0650$\,\pm\,$0.0002 \\
\ \\
B\,0747$+$426$^a$ &   $[$O{\sc ii}$]$3727 & 4483.93 &   0.2031 \\
         &    Ca{\sc ii}3934     & 4731.23 &   0.2027 \\
         &    Ca{\sc ii}3968     & 4772.31 &   0.2027 \\
         &  G-band               & 5178.74 &   0.2030 \\
         &    Mg-b               & 6228.69 &   0.2036 \\
         &                       &         &   0.2030$\,\pm\,$0.0004 \\
\ \\
B\,0750$+$434 &   $[$Ne{\sc v}$]$3426   & 4615.94 &   0.3473 \\
         &   $[$O{\sc ii}$]$3727   & 5022.40 &   0.3476 \\
         &  $[$Ne{\sc iii}$]$3869  & 5211.82 &   0.3471 \\
         &  $[$O{\sc iii}$]$4363   & 5879.17 &   0.3475 \\
         &      H$\beta$           & 6550.15 &   0.3475 \\
         &  $[$O{\sc iii}$]$4959   & 6681.97 &   0.3474 \\
         &  $[$O{\sc iii}$]$5007   & 6746.39 &   0.3474 \\
         &                         &         &   0.3474$\,\pm\,$0.0003 \\
\hline \\
\end{tabular}
\end{table}

\begin{table}[t!]
\begin{tabular}{l l l l}
\hline \\
\multicolumn{1}{c}{(1)} & \multicolumn{1}{c}{(2)} &\multicolumn{1}{c}{(3)} & \multicolumn{1}{c}{(4)}\\ 
\multicolumn{1}{c}{Source} & \multicolumn{1}{c}{Line} & \multicolumn{1}{c}{$\lambda_{\rm peak}/[$\AA$]$} & \multicolumn{1}{c}{Redshift} \\ 
\hline\\ 
\ \\
B\,0757$+$472 &      H$\gamma$        & 5019.76 &   0.1566 \\
         &      H$\beta$         & 5622.10 &   0.1566 \\
         &  $[$O{\sc iii}$]$4959 & 5736.78 &   0.1568 \\
         &  $[$O{\sc iii}$]$5007 & 5792.23 &   0.1568 \\
         &                       &         &   0.1567$\,\pm\,$0.0002 \\
\ \\
B\,0801$+$741 &      H$\beta$         & 5446.74 &   0.1205 \\
         &  $[$O{\sc iii}$]$5007 & 5610.22 &   0.1205 \\
         &    Na D               & 6603.31 &   0.1205 \\
         &    $[$O{\sc i}$]$6300 & 7058.78 &   0.1204 \\
         &      H$\alpha$        & 7351.97 &   0.1202 \\
         &   $[$S{\sc ii}$]$6583 & 7524.50 &   0.1204 \\
         &                       &         &   0.1204$\,\pm\,$0.0002 \\
\ \\
B\,0809$+$454 &   $[$O{\sc ii}$]$3727 & 4548.56 &   0.2204 \\
         &      H$\beta$         & 5933.56 &   0.2206 \\
         &  $[$O{\sc iii}$]$4959 & 6052.08 &   0.2204 \\
         &  $[$O{\sc iii}$]$5007 & 6109.84 &   0.2203 \\
         &      H$\alpha$        & 8007.07 &   0.2200 \\
         &                       &         &   0.2204$\,\pm\,$0.0003 \\
\ \\
B\,0813$+$758 &   $[$O{\sc ii}$]$3727 & 4591.00 &   0.2318 \\
         &      H$\beta$         & 5989.61 &   0.2322 \\
         &  $[$O{\sc iii}$]$4959 & 6111.66 &   0.2324 \\
         &  $[$O{\sc iii}$]$5007 & 6170.54 &   0.2324 \\
         &    $[$O{\sc i}$]$6300 & 7766.57 &   0.2328 \\
         &      H$\alpha$        & 8089.26 &   0.2326 \\
         &                       &         &   0.2324$\,\pm\,$0.0003 \\
\ \\
B\,0905$+$352 &   $[$N{\sc ii}$]$6548 & 7242.29 &   0.1059 \\
         &   $[$N{\sc ii}$]$6583 & 7279.51 &   0.1058 \\
         &                       &         &   0.106$\,\pm\,$0.001  \\
or \\
         &  G-band               & 5422.53  &  0.2596 \\
         &    Mg-b               & 6512.59  &  0.2585 \\
         &                       &          &  0.260$\,\pm\,$0.002 \\
\ \\
B\,0935$+$743 &    Ca{\sc ii}3968 & 4448.92 &   0.1212 \\
         &    Mg-b           & 5806.05 &   0.1219 \\
         &    Na D           & 6608.56 &   0.1214 \\
         &                   &         &   0.1215$\,\pm\,$0.0003 \\
\ \\ 
B\,1029$+$281 &      H$\beta$          & 5274.68 &   0.0851 \\
         &  $[$O{\sc iii}$]$4959  & 5382.18 &   0.0853 \\
         &  $[$O{\sc iii}$]$5007  & 5434.69 &   0.0854 \\
         &    Mg-b                & 5618.53 &   0.0857 \\
         &    Na D                & 6397.44 &   0.0856 \\
         &    $[$O{\sc i}$]$6300  & 6837.35 &   0.0853 \\
         &      H$\alpha$         & 7121.47 &   0.0851 \\
         &                        &         &   0.0854$\,\pm\,$0.0002 \\
\ \\
B\,1044$+$745$^a$ &    Ca{\sc ii}3968 & 4448.36 &   0.1211 \\
         &    Mg-b           & 5800.85 &   0.1209 \\
         &    Na D           & 6606.56 &   0.1211 \\
         &                   &         &   0.1210$\,\pm\,$0.0003 \\
\ \\
\hline \\
\multicolumn{4}{l}{Continued on next page...} \\
\end{tabular}
\end{table}
 
\begin{table}[t!]
\begin{tabular}{l l l l}
\multicolumn{4}{l}{...Continued from previous page} \\
\hline \\
\multicolumn{1}{c}{(1)} & \multicolumn{1}{c}{(2)} &\multicolumn{1}{c}{(3)} & \multicolumn{1}{c}{(4)}\\ 
\multicolumn{1}{c}{Source} & \multicolumn{1}{c}{Line} & \multicolumn{1}{c}{$\lambda_{\rm peak}/[$\AA$]$} & \multicolumn{1}{c}{Redshift} \\ 
\hline\\ 
B\,1110$+$405 &  G-band               & 4625.03 &   0.0743 \\
         &    Na D               & 6332.42 &   0.0746 \\
         &   $[$N{\sc ii}$]$6583 & 7073.81 &   0.0746 \\
         &                       &         &   0.0745$\,\pm\,$0.0003 \\
\ \\
B\,1213$+$422 &   $[$O{\sc ii}$]$3727  & 4632.22 &   0.2429 \\
         &  $[$Ne{\sc iii}$]$3869 & 4806.84 &   0.2424 \\
         &  $[$O{\sc iii}$]$4363  & 5422.13 &   0.2428 \\
         &      H$\beta$          & 6041.38 &   0.2428 \\
         &  $[$O{\sc iii}$]$4959  & 6162.09 &   0.2426 \\
         &  $[$O{\sc iii}$]$5007  & 6221.83 &   0.2426 \\
         &    $[$O{\sc i}$]$6300  & 7828.06 &   0.2425 \\
         &      H$\alpha$         & 8154.30 &   0.2425 \\
         &                        &         &   0.2426$\,\pm\,$0.0002 \\
\ \\
B\,1306$+$621$^a$ &    Ca{\sc ii}3934 & 4574.07 &   0.1627 \\
         &    Ca{\sc ii}3968 & 4612.79 &   0.1625 \\
         &  G-band           & 5003.63 &   0.1623 \\
         &    Mg-b           & 6016.99 &   0.1627 \\
         &                   &         &   0.1625$\,\pm\,$0.0004 \\
\ \\
B\,1310$+$451 &    Ca{\sc ii}3934     & 4075.03  &  0.0358 \\
         &    Ca{\sc ii}3968     & 4110.96  &  0.0360 \\
         &  G-band               & 4458.96  &  0.0358 \\
         &    Mg-b               & 5359.83  &  0.0357 \\
         &    Na D               & 6102.51  &  0.0356 \\
         &   $[$N{\sc ii}$]$6583 & 6817.81  &  0.0357 \\
         &                       &          &  0.0358$\,\pm\,$0.0002 \\
\ \\
B\,1416$+$380 &    Mg-b               & 5874.33 &   0.1351 \\
         &    Na D               & 6688.00 &   0.1349 \\
         &    $[$O{\sc i}$]$6300 & 7150.68 &   0.1350 \\
         &      H$\alpha$        & 7448.93 &   0.1350 \\
         &   $[$N{\sc ii}$]$6583 & 7471.70 &   0.1350 \\
         &                       &         &   0.1350$\,\pm\,$0.0002 \\
\ \\
B\,1426$+$295 &  G-band               & 4678.63 &   0.0868 \\
         &  $[$O{\sc iii}$]$4363 & 4742.83 &   0.0871 \\
         &      H$\beta$         & 5283.63 &   0.0869 \\
         &    Mg-b               & 5627.45 &   0.0874 \\
         &    Na D               & 6403.01 &   0.0865 \\
         &   $[$N{\sc ii}$]$6583 & 7156.50 &   0.0871 \\
         &                       &         &   0.0870$\,\pm\,$0.0003 \\
\ \\
B\,1543$+$845$^a$ &  $[$O{\sc iii}$]$4959 & 5955.22 &   0.2009 \\
         &  $[$O{\sc iii}$]$5007 & 6013.78 &   0.2011 \\
         &                       &         &   0.201$\,\pm\,$0.001  \\
\ \\
B\,1709$+$464 &    Ca{\sc ii}3968     & 4114.19 &   0.0368 \\
         &  G-band               & 4463.67 &   0.0369 \\
         &    Mg-b               & 5366.13 &   0.0369 \\
         &    Na D               & 6108.30 &   0.0365 \\
         &   $[$N{\sc ii}$]$6583 & 6824.31 &   0.0367 \\
         &                       &         &   0.0368$\,\pm\,$0.0002 \\
\ \\
B\,1736$+$375 &    Ca{\sc ii}3968 & 4548.75 &   0.1563 \\
         &    Mg-b           & 5984.90 &   0.1565 \\
         &    Na D           & 6811.76 &   0.1559 \\
         &                   &         &   0.1562$\,\pm\,$0.0003 \\
\hline \\
\end{tabular}
\end{table}
 
\begin{table}[t!]
\begin{tabular}{l l l l}
\multicolumn{4}{l}{\ } \\
\hline \\
\multicolumn{1}{c}{(1)} & \multicolumn{1}{c}{(2)} &\multicolumn{1}{c}{(3)} & \multicolumn{1}{c}{(4)}\\ 
\multicolumn{1}{c}{Source} & \multicolumn{1}{c}{Line} & \multicolumn{1}{c}{$\lambda_{\rm peak}/[$\AA$]$} & \multicolumn{1}{c}{Redshift} \\ 
\hline\\ 
B\,1852$+$507 &  G-band & 4717.29 &   0.0958 \\
         &    Mg-b & 5670.70 &   0.0958 \\
         &    Na D & 6457.09 &   0.0957 \\
         &         &         &   0.0958$\,\pm\,$0.0003 \\
\ \\
B\,1911$+$470 &  G-band               & 4540.16 &   0.0546 \\
         &    Mg-b               & 5457.87 &   0.0547 \\
         &    Na D               & 6216.32 &   0.0549 \\
         &   $[$N{\sc ii}$]$6583 & 6944.73 &   0.0549 \\
         &                       &         &   0.0548$\,\pm\,$0.0002 \\
\ \\
B\,1918$+$512 &    Ca{\sc ii}3968     & 5050.42 &   0.2838 \\
         &  $[$O{\sc iii}$]$5007 & 6425.44 &   0.2833 \\
         &                       &         &   0.284$\,\pm\,$0.001\\
\ \\
B\,2147$+$816 &   $[$O{\sc ii}$]$3727  & 4270.15 &   0.1457 \\
         &  $[$Ne{\sc iii}$]$3869 & 4432.08 &   0.1455 \\
         &      H$\gamma$         & 4972.15 &   0.1457 \\
         &  $[$O{\sc iii}$]$4363  & 4998.67 &   0.1457 \\
         &      H$\beta$          & 5568.67 &   0.1456 \\
         &  $[$O{\sc iii}$]$4959  & 5681.59 &   0.1457 \\
         &  $[$O{\sc iii}$]$5007  & 5736.42 &   0.1457 \\
         &                        &         &   0.1457$\,\pm\,$0.0001 \\
\hline\\
\end{tabular}
\ \\
Notes:\\
$a-$Spectrum has been smoothed with a 3 pixel rectangular box.\\
\vspace*{\textheight}
\end{table}

\clearpage

\begin{table*}
\centering
\setlength{\tabcolsep}{12pt}
\caption{\label{DS1923:figure_captions} 
The flux density levels of the first contour in the presented radio contour maps. Column 1 gives the name of the source; columns 2 to 5 give the flux density of the first contour in the radio contour plots of the WENSS, NVSS, FIRST and WSRT radio map. Column 6 gives the beam size of the WSRT beam in cases where a WSRT radio map is presented. Column 7 gives the origin of the presented optical images. Here, PI stands for POSS-I, PII for POSS-II; LDSS is an imaging spectrograph on the 4.2-m WHT on La Palma, HARIS is an imaging spectrograph on the 2.2-m University of Hawaii telescope on Mauna Kea. Unless indicated otherwise, contour levels have been plotted at $-1,1,2,4,8,16,32,64,128,256$ times the flux density level presented in this table.}
\begin{tabular}{l c l l l l@{$\,\times\,$}r l}  
\hline \hline \\
\multicolumn{1}{c}{(1)} & \multicolumn{1}{c}{(2)} &  \multicolumn{1}{c}{(3)} & \multicolumn{1}{c}{(4)} & \multicolumn{1}{c}{(5)} &\multicolumn{2}{c}{(6)} & \multicolumn{1}{c}{(7)}\\ 
Source   & \multicolumn{1}{c}{WENSS} & \multicolumn{1}{c}{NVSS} & \multicolumn{1}{c}{FIRST} & \multicolumn{1}{c}{WSRT} & \multicolumn{2}{c}{Beam size} & \multicolumn{1}{c}{Optical} \\
         & \multicolumn{4}{c}{\hrulefill~$[$mJy beam$^{-1}\,]$~\hrulefill}& \multicolumn{2}{c}{$[\,\arcsec\,\times\,\arcsec\,]$}& \\[1ex]
\hline \\ 
B\,0211$+$326 & 8  &     &      & 2.5 & 26.11 & 12.56          & PII \\
B\,0217$+$367 & 8  & 1.2 &      & 1.2 & 20.85 & 12.65          & PII \\
B\,0648$+$733 & 10 & 1.3 &      & 0.8 & 17.06 & 11.18          & PI  \\
B\,0648$+$431 & 8  & 1.3 &      & 0.8 & 20.85 & 12.65          & PII \\
B\,0658$+$490 & 8  & 1.3 & 0.55 &     & \multicolumn{2}{c}{\ } & PII \\   
B\,0747$+$426 & 8  & 1.5 & 0.45 &     & \multicolumn{2}{c}{\ } & PII \\
B\,0750$+$434 & 8  & 1.3 & 0.45 &     & \multicolumn{2}{c}{\ } & PII \\
B\,0757$+$477 & 10 & 1.3 & 0.45 &     & \multicolumn{2}{c}{\ } & PII \\
B\,0801$+$741 & 11 & 1.3 &      & 0.8 & 16.39 & 11.30          & PII \\ 
B\,0809$+$454 & 10 & 1.3 & 0.45 &     & \multicolumn{2}{c}{\ } & PII \\
B\,0813$+$758 & 8  & 1.3 &      & 1.5 & 21.80 & 9.70           & PII \\
B\,0905$+$352 & 8  &     & 0.45 &     & \multicolumn{2}{c}{\ } & PI  \\
B\,0935$+$743 & 10 & 1.2 &      &     & \multicolumn{2}{c}{\ } & PII \\
B\,1029$+$281 & 18 & 1.2 & 0.55 &     & \multicolumn{2}{c}{\ } & PII \\ 
B\,1044$+$745 & 7$^a$ & 1.2$^a$ &  &  & \multicolumn{2}{c}{\ } & PI  \\
B\,1110$+$405 & 8  & 1.3 & 0.70 &     & \multicolumn{2}{c}{\ } & PII \\
B\,1213$+$422 & 8  &     & 0.45 &     & \multicolumn{2}{c}{\ } & PII \\
B\,1245$+$676 & 8  & 1.2 &      &     & \multicolumn{2}{c}{\ } & PI  \\
B\,1306$+$621 & 8  & 1.2 &      & 0.5 & 16.63 & 12.79          & PI  \\
B\,1310$+$451 & 8  & 1.3 & 0.50 &     & \multicolumn{2}{c}{\ } & PI  \\ 
B\,1416$+$380 & 8  & 1.3 & 0.45 &     & \multicolumn{2}{c}{\ } & PII \\ 
B\,1426$+$295 & 12 & 1.5 & 0.60 &     & \multicolumn{2}{c}{\ } & PII \\
B\,1543$+$845 & 7  & 1.3 &      &     & \multicolumn{2}{c}{\ } & HARIS\\
B\,1709$+$464 & 10 & 1.3 & 0.45 &     & \multicolumn{2}{c}{\ } & PII \\
B\,1736$+$375 & 8  & 1.3 &      &     & \multicolumn{2}{c}{\ } & PI  \\
B\,1852$+$507 & 8  & 1.3 &      &     & \multicolumn{2}{c}{\ } & PII \\
B\,1911$+$470 & 8  & 1.3 &      &     & \multicolumn{2}{c}{\ } & PII \\
B\,1918$+$516 & 8  &     &      & 0.5 & 16.22 & 12.25          & LDSS \\
B\,2147$+$816 & 5  & 1.3 &      &     & \multicolumn{2}{c}{\ } & PII \\
\hline \\
\end{tabular}
\ \\
\begin{minipage}{0.8\linewidth}
Notes:\\
$a-$Extra contour plotted at a level of $\sqrt2$ times the lowest contour level.\\
\end{minipage}
\vspace*{\textheight}
\end{table*}

\newpage

\begin{figure*}
\centering
\vfill
\ \\
\caption{\label{DS1923:0211+326}
B\,0211+326: The WENSS radio contour plot, an overlay of the WSRT radio map (contours) with an optical image (grey scale; the identification has been encircled) and the optical spectrum of the host galaxy.}  
\end{figure*}
\newpage
\begin{figure*}
\centering
\caption{\label{DS1923:0217+367}
B\,0217+367: The WENSS (upper) and NVSS (lower) radio contour plots, an overlay of the WSRT radio map (contours) with an optical image (grey scale) and the optical spectrum of the host galaxy.}  
\end{figure*}

\begin{figure*}
\centering
\caption{\label{DS1923:0648+733}
B\,0648+733: The WENSS (left) and NVSS (right) radio contour plots, an overlay of the WSRT radio map (contours) with an optical image (grey scale) and the optical spectrum of the host galaxy.}
\end{figure*}

\begin{figure*}
\centering
\caption{\label{DS1923:0648+431}
B\,0648+431: The WENSS (upper) and NVSS (lower) radio contour plots, an overlay of the WSRT radio map (contours) with an optical image (grey scale) and the optical spectrum of the host galaxy.}
\end{figure*}

\begin{figure*}
\centering
\caption{\label{DS1923:0658+490}
B\,0658+490: The WENSS (upper) and NVSS (lower) radio contour plots, an overlay of the WSRT radio map (contours) with an optical image (grey scale) and the optical spectrum of the host galaxy.}
\end{figure*}

\begin{figure*}
\centering
\caption{\label{DS1923:0747+426}
B\,0747+426: The WENSS (upper) and NVSS (lower) radio contour plots, an overlay of the FIRST radio map (contours) with an optical image (grey scale; the identification has been encircled) and the optical spectrum of the host galaxy.}
\end{figure*}

\begin{figure*}
\centering
\caption{\label{DS1923:0750+434}
B\,0750+434: The WENSS (upper) and NVSS (lower) radio contour plots, an overlay of the FIRST radio map (contours) with an optical image (grey scale) and the optical spectrum of the host galaxy.}
\end{figure*}

\begin{figure*}
\centering
\caption{\label{DS1923:0757+477}
B\,0757+477: The WENSS (upper) and NVSS (lower) radio contour plots, an overlay of the FIRST radio map (contours) with an optical image (grey scale) and the optical spectrum of the host galaxy.} 
\end{figure*}

\begin{figure*}
\centering
\caption{\label{DS1923:0801+741}
B\,0801+741: The WENSS (upper) and NVSS (lower) radio contour plots, an overlay of the WSRT radio map (contours) with an optical image (grey scale) and the optical spectrum of the host galaxy.}
\end{figure*}

\begin{figure*}
\centering
\caption{\label{DS1923:0809+454}
B\,0809+454: The WENSS (upper) and NVSS (lower) radio contour plots, an overlay of the FIRST radio map (contours) with an optical image (grey scale) and the optical spectrum of the host galaxy.}
\end{figure*}

\begin{figure*}
\centering
\caption{\label{DS1923:0813+758}
B\,0813+758: The WENSS (upper) and NVSS (lower) radio contour plots, an overlay of the WSRT radio map (contours) with an optical image (grey scale) and the optical spectrum of the host galaxy.}
\end{figure*}

\begin{figure*}
\centering
\caption{\label{DS1923:0905+352}
B\,0905+352: The WENSS radio contour plot, an overlay of the FIRST radio map (contours) with an optical image (grey scale; the identification has been encircled) and the optical spectrum of the host galaxy.}
\end{figure*}

\begin{figure*}
\centering
\caption{\label{DS1923:0935+743}
B\,0935+352: The WENSS radio contour plot, an overlay of the NVSS radio map (contours) with an optical image (grey scale) and the optical spectrum of the host galaxy.}
\end{figure*}

\begin{figure*}
\centering
\caption{\label{DS1923:1029+281}
B\,1029+281: The WENSS (upper) and NVSS (lower) radio contour plots, an overlay of the FIRST radio map (contours) with an optical image (grey scale) and the optical spectrum of the host galaxy.}
\end{figure*}

\clearpage

\begin{figure*}
\centering
\caption{\label{DS1923:1044+745}
B\,1044+745: The WENSS radio contour plot, an overlay of the NVSS radio map (contours) with an optical image (grey scale) and the optical spectrum of the host galaxy.} 
\end{figure*}

\begin{figure*}
\centering
\caption{\label{DS1923:1110+405}
B\,1110+405: The WENSS (upper) and NVSS (lower) radio contour plots, an overlay of the FIRST radio map (contours) with an optical image (grey scale) and the optical spectrum of the host galaxy.}
\end{figure*}

\begin{figure*}
\centering
\caption{\label{DS1923:1213+422}
B\,1213+422: The WENSS radio contour plot, an overlay of the FIRST radio map (contours) with an optical image (grey scale) and the optical spectrum of the host galaxy.}
\end{figure*}

\begin{figure*}
\centering
\caption{\label{DS1923:1245+676}
B\,1245+676: The WENSS radio contour plot and an overlay of the NVSS radio map (contours) with an optical image (grey scale). An optical spectrum has been published by Marcha et al. (\cite{marcha96}).}
\end{figure*}

\begin{figure*}
\centering
\caption{\label{DS1923:1306+621}
B\,1306+621: The WENSS (left) and NVSS (right) radio contour plots, an overlay of the WSRT radio map (contours) with an optical image (grey scale) and the optical spectrum of the host galaxy.}
\end{figure*}

\begin{figure*}
\centering
\caption{\label{DS1923:1310+451}
B\,1310+451: The WENSS (upper) and NVSS (lower) radio contour plots, an overlay of the FIRST radio map (contours) with an optical image (grey scale) and the optical spectrum of the host galaxy.} 
\end{figure*}

\begin{figure*}
\centering
\caption{\label{DS1923:1416+380}
B\,1416+380: The WENSS (upper) and NVSS (lower) radio contour plots, an overlay of the FIRST radio map (contours) with an optical image (grey scale) and the optical spectrum of the host galaxy.}
\end{figure*}

\begin{figure*}
\centering
\caption{\label{DS1923:1426+295}
B\,1426+295: The WENSS (left) and NVSS (right) radio contour plots, an overlay of the FIRST radio map (contours) with an optical image (grey scale) and the optical spectrum of the host galaxy.}
\end{figure*}

\begin{figure*}
\centering
\caption{\label{DS1923:1543+845}
B\,1543+845: The WENSS (upper) and NVSS (lower) radio contour plots, an overlay of the NVSS radio map (contours) with an optical image (grey scale) and the optical spectrum of the host galaxy.}
\end{figure*}

\begin{figure*}
\centering
\caption{\label{DS1923:1709+464}
B\,1709+464: The WENSS (upper) and NVSS (lower) radio contour plots, an overlay of the FIRST radio map (contours) with an optical image (grey scale) and the optical spectrum of the host galaxy.} 
\end{figure*}

\begin{figure*}
\centering
\caption{\label{DS1923:1736+375}
B\,1736+375: The WENSS radio contour plot, an overlay of the NVSS radio map (contours) with an optical image (grey scale) and the optical spectrum of the host galaxy of the central source; note the fuzzy optical object near the center of the northern radio component.}
\end{figure*}

\begin{figure*}
\centering
\caption{\label{DS1923:1852+507} 
B\,1852+507: The WENSS radio contour plot, an overlay of the NVSS radio map (contours) with an optical image (grey scale) and the optical spectrum of the host galaxy.}
\end{figure*}

\clearpage

\begin{figure*}
\centering
\caption{\label{DS1923:1911+470}
B\,1911+470: The WENSS radio contour plot, an overlay of the NVSS radio map (contours) with an optical image (grey scale) and the optical spectrum of the host galaxy.}
\end{figure*}

\begin{figure*}
\centering
\caption{\label{DS1923:1918+516}
B\,1918+516: The WENSS radio contour plot, an overlay of the 1.4-GHz WSRT radio map (contours) with an optical image (grey scale; the identification is indicated by the arrow) and the optical spectrum of the host galaxy.}
\end{figure*}

\begin{figure*}
\centering
\caption{\label{DS1923:2147+816}
B\,2147+816: The WENSS (upper) and NVSS (lower) radio contour plots, an overlay of the NVSS radio map (contours) with an optical image (grey scale) and the optical spectrum of the host galaxy.}
\end{figure*}

\end{appendix}

\begin{thebibliography}{}

\bibitem[1989]{barthel89} Barthel, P.D. 1989, ApJ, 336, 606
\bibitem[1995]{becker95} Becker, R., White, R., Helfand, D. 1995, ApJ, 450, 559
\bibitem[1998]{bhatnagar98} Bhatnagar, S., Gopal-Krishna, Wisotzki, L. 1998, MNRAS, 299, L25
\bibitem[1999]{blundell99} Blundell, K., Rawlings, S., Willott, C.J. 1999, AJ, 117, 677 
\bibitem[1972]{burbidge72} Burbidge, E.M., Strittmattar, P.A. 1972, ApJ, 172, L37 
\bibitem[1981]{breugel81} van Breugel, W.J.M., Willis, A.G. 1981, A\&A, 96, 332 
\bibitem[1982]{breugel82} van Breugel, W.J.M., J\"{a}gers W. 1982, A\&AS, 49, 529 
\bibitem[1989]{debruyn89} de Bruyn, A.G. 1989, A\&A, 226, L13 
\bibitem[1975]{colla75} Colla, G., Fanti, C., Fanti, R., et al. 1975, A\&AS, 20, 1
\bibitem[1998]{condon98} Condon, J.J., Cotton, W.D., Greisen, E.W., et al. 1998, AJ, 115, 1693
\bibitem[1970]{demoulin70} Demoulin, M. 1970, ApJ, 160, L79 
\bibitem[1990]{dingley90} Dingley, S.J. 1990, Ph.D.-Thesis, University of Cambridge
\bibitem[1990]{djorgovski90} Djorgovski, S.G., Thompson, D., Vigotti, M., Grueff, G. 1990, PASP, 102, 113
\bibitem[1995]{djorgovski95} Djorgovski, S.G., Thompson, D., Maxfield, L.,
Vigotti, M., Grueff, G. 1995, ApJS, 101, 255 
\bibitem[1991]{falle91} Falle, S.A.E.G. 1991, MNRAS, 250, 581
\bibitem[1974]{fanaroff74} Fanaroff, B.L., Riley, J.M. 1974, MNRAS, 167, 31
\bibitem[1985]{faulkner85} Faulkner, M.A 1985, Ph.D.-Thesis, Univ. of Cambridge
\bibitem[1978]{fomalont78} Fomalont, E.B., Bridle, A.H. 1978, AJ, 83, 7 
\bibitem[1992]{degrijp92} de Grijp, M.H.K., Keel, W.C., Miley, G.K., Goudfrooij, P., Lub, J. 1992, A\&AS, 96, 389
\bibitem[1979]{hine79} Hine, R.G., 1979, MNRAS, 189, 527 
\bibitem[1999]{ishwara-chandra99} Ishwara-Chandra, C.H., Saikia, D.J. 1999, MNRAS, 309, 100
\bibitem[1986]{jagers86} J\"{a}gers, W.J. 1986, Ph.D.-thesis, University of Leiden
\bibitem[1999]{kaiser99} Kaiser, C.R., Alexander, P. 1999, MNRAS, 302, 515 
\bibitem[1997]{kaiser97} Kaiser, C.R., Dennett-Thorpe, J., Alexander,
P. 1997, MNRAS, 292, 723
\bibitem[1993]{lacy93} Lacy, M., Rawlings, S., Saunders, R., Warner,
P.J. 1993, MNRAS, 264, 721
\bibitem[1983]{laing83} Laing, R.A., Riley, J.M., Longair, M.S. 1983, MNRAS, 204, 151  
\bibitem[2001]{lara01} Lara, L., Cotton, W.D., Feretti, L., et al. 2001, A\&A, {\it in press}
\bibitem[1997]{mack97} Mack, K.-H., Klein, U., O'Dea, C.P., Willis, A.G. 1997, A\&AS, 123, 423
\bibitem[1998]{mack98} Mack, K.-H., Klein, U., O'Dea, C.P., Willis, A.G., Saripalli, L. 1998, A\&A, 329, 431
\bibitem[1996]{marcha96} Marcha, M.J.M., Browne, I.W.A., Impey, C.D., Smith, P.S. 1996, MNRAS, 281, 425
\bibitem[1979]{masson79} Masson, C.R. 1979, MNRAS, 187, 253
\bibitem[1979]{miley79} Miley, G.K., Osterbrock, D.E. 1979, PASP, 92, 257
\bibitem[2000]{palma00} Palma, C., Bauer, F.E., Cotton, et al. 2000, AJ, 119, 2068  
\bibitem[1996]{parma96} Parma, P., de Ruiter, H.R., Mack, K.-H., et al. 1996, A\&A 306, 708 
\bibitem[1999]{parma99} Parma, P., Murgia, M., Morganti, R., et al. 1999, A\&A, 344, 7
\bibitem[1984]{perley84} Perley, R.A., Bridle, A.H., Willis, A.G. 1984, ApJS, 54, 291 
\bibitem[1997]{rengelink97} Rengelink, R., Tang, Y., de Bruyn, A.G., et al. 1997, A\&AS, 124,259 
\bibitem[1989]{riley89} Riley, J.M. 1989, MNRAS, 238, 1055
\bibitem[1988]{riley88} Riley, J.M., Warner, P.J., Rawlings, S., et al. 1988, MNRAS, 236, 13 
\bibitem[1996]{rottgering96} R\"{o}ttgering, H.J.A., Tang, Y., Bremer,
M.A.R., et al. 1996, MNRAS, 282, 1033  
\bibitem[1973]{sargent73} Sargent, W.L.W. 1973, ApJ, 182, L13 
\bibitem[1982]{saunders82} Saunders, R.D.E. 1982, Ph.D.-Thesis, University of Cambridge 
\bibitem[1987]{saunders87} Saunders, R.D.E., Baldwin, J.E., Warner, P.J. 1987, MNRAS, 225, 713
\bibitem[1986]{saripalli86} Saripalli, L., Gopal-Krishna, Reich, W., K\"{u}hr, H. 1986, A\&A, 170, 20
\bibitem[1974]{scheuer74} Scheuer, P.A.G. 1974, MNRAS, 166, 513
\bibitem[1999a]{schoenmakers99a} Schoenmakers, A.P. 1999, Ph.D.-Thesis, Utrecht University
\bibitem[1999b]{schoenmakers99b} Schoenmakers, A.P., de Bruyn, A.G.,
R\"{o}ttgering, H.J.A., van der Laan, H. 1999, A\&A, 341, 44
\bibitem[2000a]{schoenmakers00a} Schoenmakers, A.P., de Bruyn, A.G.,
R\"{o}ttgering, H.J.A., van der Laan, H., Kaiser, C.R. 2000a, MNRAS, 315, 371
\bibitem[2000b]{schoenmakers00b} Schoenmakers, A.P., Mack, K.-H., de Bruyn,
A.G., et al. 2000b, A\&AS, 146, 293
\bibitem[1997]{simien97} Simien, F., Prugniel, P. 1997, A\&AS, 122, 521 
\bibitem[1985]{spinrad85} Spinrad, H., Djorgovski, S., Marr, J., Aguilar, L. 1985, PASP, 97, 932
\bibitem[1980]{strom80} Strom, R.G., Willis, A.G. 1980, A\&A, 85, 36 
\bibitem[1981]{strom81} Strom, R.G., Baker J.R., Willis, A.G. 1981, A\&A, 100, 220 
\bibitem[1983]{strom83} Strom, R.G., Fanti, R., Parma, P., Ekers, R.D. 1983, A\&A, 122, 305
\bibitem[1993]{subrahmanyan93} Subrahmanyan, R., Saripalli, L. 1993, MNRAS, 260, 908
\bibitem[1989]{vigotti89} Vigotti, M., Grueff, G., Perley, R., Clark, B.G., Bridle, A.H. 1989, AJ, 98, 419  
\bibitem[1977]{wagett77} Wagett, P.C., Warner, P.J., Baldwin, J.E. 1977, MNRAS, 181, 465 
\bibitem[1991]{wieringa91} Wieringa, M., 1991, Ph.D.-Thesis, University of Leiden
\bibitem[1974]{willis74} Willis, A.G., Strom, R.G., Wilson, A.S. 1974, Nat., 260, 625
\bibitem[1981]{willis81} Willis, A.G., Strom, R.G., Bridle, A.H., Fomalont, E.B. 1981, A\&A, 95, 250 
\end{thebibliography}
\end{document}